\documentstyle[12pt,righttag,amssymb]{amsart} 

\newtheorem{thm}{Theorem}[section]
\newtheorem{cor}[thm]{Corollary}
\newtheorem{lem}[thm]{Lemma}
\newtheorem{prop}[thm]{Proposition}

\theoremstyle{definition}
\newtheorem{defn}{Definition}[section]
\newtheorem{acknowledgement}{Acknowledgements}

\theoremstyle{remark}
\newtheorem{rem}{Remark}[section]
\newtheorem{example}{Example}[section]

\numberwithin{equation}{section}

\newcommand{\thmref}[1]{Theorem~\ref{#1}}
\newcommand{\secref}[1]{Section~\ref{#1}}
\newcommand{\lemref}[1]{Lemma~\ref{#1}}
\newcommand{\propref}[1]{Proposition~\ref{#1}}

\renewcommand{\rm}{\fontshape{n}\selectfont}
\newcommand{\ncite}[1]{[#1]}
\newcommand{\Alt}{\operatornamewithlimits{Alt}}
\newcommand{\nquad}{{\!\!\!\!\!\!}}

\newcommand{\nqqquad}{\nquad\nquad\nquad}
\newcommand{\id}{\operatorname{id}}
\newcommand{\ad}{\operatorname{ad}}
\newcommand{\Vir}{\operatorname{Vir}}
\newcommand{\tens}{\otimes}
\newcommand{\<}{\langle}
\renewcommand{\>}{\rangle}
\newcommand{\be}{\begin{equation}}
\newcommand{\dq}{\partial_q}
\newcommand{\binq}[2]{\left(\matrix #1\cr#2 \endmatrix\right)_q}

\long\def\bullshit#1\endbullshit{\par\noindent\nqqquad\nqqquad
{\bf{\size{14}{16pt}\selectfont Bullshit?}\par #1 
\par\noindent\nqqquad\nqqquad{\size{14}{16pt}\selectfont Or no?}}\par}

\def\hfl#1#2{\smash{\mathop{\hbox to 6mm{\rightarrowfill}}
\limits^{\scriptstyle#1}_{\scriptstyle#2}}}
\def\hfll#1#2{\smash{\mathop{\hbox to 6mm{\leftarrowfill}}
\limits^{\scriptstyle#1}_{\scriptstyle#2}}}
\def\vfl#1#2{\llap{$\scriptstyle #1$}\left\downarrow
\vbox to 3mm{}\right.\rlap{$\scriptstyle #2$}}

\def\build#1_#2^#3{\mathrel{
\mathop{\kern 0pt#1}\limits_{#2}^{#3}}}

\def\wtilde {\widetilde}

\let\nn=\noindent

\let\noi=\noindent

\let\what=\widehat

\newcommand{\Diff}{\operatorname{Diff}}
\newcommand{\Tr}{\operatorname{Tr}}
\newcommand{\tr}{\operatorname{tr}}
\newcommand{\Der}{\operatorname{Der}}
\def\un{1}
\def\lbc{\lbrace}
\def\rbc{\rbrace}
\def\lbk{\lbrack}
\def\rbk{\rbrack}
\def\Ja{{\cal J}}

\def\G{{\frak G}}

\def\gg{{\frak g}}
\def\A{{\frak A}}
\def\F{{\cal F}}

\def\CC{{\Bbb C}}
\def\C{{\Bbb C}}

\def\RR{{\Bbb R}}
\def\R{{\Bbb R}}

\def\TT{{\Bbb T}}
\def\ZZ{{\Bbb Z}}
\def\Z{{\Bbb Z}}

\def\a{\alpha}
\def\b{\beta}
\def\d{\delta}

\def\eps{\epsilon}
\def\g{\gamma}

\def\l{\lambda}
\def\om{\omega}

\def\s{\sigma}
\def\t{\theta}

\def\ve{\varepsilon}

\def\hh{\widehat}

\newcommand{\res}{\operatorname{res}}
\def\ra{\rightarrow}
\def\sbs{\subset}
\def\la{\leftarrow}
\def\part{\partial}
\def\Inf{\infty}

\def\op{\oplus}
\def\ot{\otimes}

\def\bgt{\nabla}
\def\mps{\mapsto}
\def\ts{\times}

\newcommand{\Lie}{\operatorname{Lie}}

\newcommand{\tenfm}{\frak}
\newcommand{\got}{\frak}

\begin{document}

\title[$q$-pseudodifferential symbols]{Extensions 
and contractions of the Lie algebra of $q$-pseudodifferential symbols
on the circle}

\author[B. Khesin]{Boris KHESIN\mbox{${}^{\protect\dag}$}}
\address[B. Khesin]
{Department of Mathematics\\ Yale University\\ New Haven, CT 06520\\ USA}
\email{khesin@@math.yale.edu} 
\thanks{\mbox{${}^{\protect\dag}$}Research was supported in 
part by the NSF grant DMS-9307086.}

\author[V. Lyubashenko]{Volodymyr LYUBASHENKO\mbox{${}^{\protect\ddag}$}}
\address[V. Lyubashenko]{Department of Mathematics\\ University of York\\ 
Heslington, York, YO1 5DD\\ England, UK}
\email{vvl1@@unix.york.ac.uk}
\thanks{\mbox{${}^{\protect\ddag}$}Research was supported in 
part by the EPSRC research grant GR/G 42976.}

\author[C. Roger]{Claude ROGER}
\address[C. Roger]
{Laboratoire de Geometrie et Analyse, URA 746\\ Universite Lyon-I\\ 
43, Blvd. du 11 Novembre 1918, Bat.101\\ 69622 Villeurbanne Cedex\\ FRANCE}
\email{roger@@geometrie.univ-lyon1.fr}      

\date {\today ; June 7, 1995}

\begin{abstract}
We construct cocycles on the Lie algebra of pseudo- and
$q$-pseudodif\-fe\-ren\-tial symbols of one variable and on their close 
relatives: the sine-algebra and the Poisson algebra on two-torus. 
A ``quantum'' Godbillon-Vey cocycle on (pseudo)-differential operators 
appears in this construction as a natural generalization of the 
Gelfand-Fuchs 3-cocycle on periodic vector fields. We describe a 
nontrivial embedding of the Virasoro algebra into (a completion of) 
$q$-pseudodifferential symbols, and propose $q$-analogs of the KP and 
KdV-hierarchies admitting an infinite number of conserved charges.
\end{abstract}

\maketitle

\section{Introduction} 
The Lie algebra of symbols of pseudodifferential operators on the circle 
arises naturally in various contexts: it is  a deformation of the Poisson 
algebra of functions on the two-dimensional torus, the phase space 
for the KP and KdV hierarchies, a generalization of the Virasoro algebra 
and of the Lie algebra of differential operators. The latter object 
appeared recently in conformal field theory under
the name of  $W_\Inf$ algebra.

We start our consideration with the Poisson algebra $C^\Inf({ T^2 })$ 
of smooth functions on a two-dimensional torus (with respect to the 
standard Poisson bracket on $T^2$). Various deformation of this
object include the sine-algebras \cite{[H-O-T]}, algebras of differential and
pseudo\-differential operators \cite{[P-S]}, $q$-analogs of differential 
operators and (finite-dimensional) $gl(n)$-approximations. 

In this paper we consider so called $q$-deformation of the Lie algebra of
pseudodifferential operators ($\psi DO_q$) on the circle. 
From the quantization point 
of view we consider in this paper the next link in the chain of 
deformations of the Poisson algebra to pseudodifferential
 operators and then to $q$-pseudodifferential operators.
Our work is  closely connected with Kassel's paper \cite{[Ka]} where 
$q$-analogs of the algebra of purely differential operators and of
the corresponding cocycles were described.

The construction can be applied to the case of pseudodifferential
operators in $\C^n$, where one has $2n$ outer derivations coming from
commutations with $\log D_i$ and $\log x_i$ (see e.g. Dhumaduldaev~
\cite{Dzha}). However, the Poisson structures related to integrable 
systems single out the one-dimensional case.

Some our results were generalised by Wambst \cite{Wam:pDO}, who
calculated, in particular, the first cyclic cohomology group of
the topological algebra of $q$-pseudodifferential symbols 
and found that our cocycles span linearly the whole group.

We propose a $q$-version of the Manin triple of differential and integral
operators on the circle. The extension of this triple by the
logarithm of the derivative operator (\cite{[K-K]}) and by the corresponding 
central term was considered in detail in \cite{[K-Z]}. In the classical
situation the corresponding (infinite dimensional) Lie-Poisson group is 
``unipotent'' and equipped with the quadratic generalized Gelfand-Dickey 
Poisson structure. This structure is the main ingredient in the 
construction of the KP and $n$-KdV hierarchies (\cite{[G-D],[K-Z]}).

We show that the group of $q$-pseudodifferential operators is rather
``solvable'' than ``unipotent''. Its  exponential map is still well defined
and surjective as in the classical case, but is not bijective. This
allows to define $q$-analogs of the hierarchies which are now commuting 
flows having infinite number of conserved quantities on the space of
$q$-pseudodifferential operators.

This paper is organized as follows.
In the second Section we discuss the Poisson algebra on a torus and its 
approximations by finite dimensional algebras $u((m+1)/2,(m-1)/2)$.
 Section~3 is devoted to the description of various relations between 
the algebra of pseudodifferential operators, its $q$-analog, and the 
sine-algebra. 
In Section~4 we define a 
``$q$-logarithmic'' outer derivation of the algebra and the 
corresponding cyclic 2-cocycle and (the Lie algebra 2-cocycle). 
We construct also quantum Godbillon--Vey cocycle and higher degree
generalizations. They are defined for any $q\ne0$ and depend on $q$
continuously.
In Section~5 we study the behaviour of cocycles when the algebra is 
deformed. An embedding of the undeformed Virasoro algebra into an 
appropriately completed algebra of $q$-pseudodifferential operators 
is constructed in Section~6. The final Section~7 contains the construction 
of $q$-analogs of the KP and $n$-KdV hierarchies inspired by the previous
formalism.

\begin{acknowledgement}
We are pleased to thank M.~I. Golenishcheva-Kutuzova, A.~N. Kirillov, 
M.~V. Saveliev and A.~M. Vershik for fruitful discussions. 
We also thank V.~G. Kac and A.~O. Radul for 
communicating us preliminary version of their work~\cite{[K-R]}.
\end{acknowledgement}

\section{Deformations of the Poisson algebra on two-torus}
Let $\TT^2$ be the standard torus with $p, q$-coordinates equipped with a
symplectic form $\om = dp \wedge dq$, and 
$N = C^\Inf \left({ \TT^2 }\right)$ be the Lie algebra of (periodic) 
functions in $p, q$ with respect to the Poisson bracket.

\subsection{Low-dimensional Lie cocycles on $C^\Inf \left({ \TT^2 }\right)$}
\begin{thm} [see, e.g. \cite{[L-R1]}, \cite{[Ki]}]
 The group 
$H_{\Lie}^2 \left({ C^\Inf \left({ \TT^2 }\right), \RR }\right)$ is
two-dimen\-si\-o\-nal and is generated by the 2-cocycles
$$c_p (f, g) =
\int_{ }^{ }\!\!\int_{\TT^2 }^{ }
\left(\{p, f \} g \right)\om = 
\int_{ }^{ }\!\!\int_{\TT^2 }^{ }
\left({ { \part f \over \part q}\cdot g }\right)\om$$
$$c_q (f, g) =
\int_{ }^{ }\!\!\int_{\TT^2 }^{ }
\left(\{q, f \} g \right)\om = -
\int_{ }^{ }\!\!\int_{\TT^2 }^{ }
\left({ { \part f \over \part p}  \cdot g }\right)\om,$$
where $ f, g \in N = 
C^\Inf \left({ \TT^2 }\right).$
\end{thm}

\begin{rem}
The linear functions $p$ and $q$ are being considered as multivalued 
function on $\TT^2$. They define Hamiltonian vector fields which ``shift 
the mass center'' of $\TT^2$ unlike the fields corresponding to 
univalued Hamiltonians.
\end{rem}

More generally (see \cite{[L-R1]}) on an arbitrary compact symplectic manifold
$(M,\om)$ the Poisson algebra of functions 
$C^\Inf (M)$ has the universal central extension of  dimension 
$b_1 = \dim H_1 (M, \RR)$
(i.e. $H_{\Lie}^2
\left({ C^\Inf (M), \RR }\right) \simeq H_1 (M, \RR)$).

\begin{thm}\label{c0c1}
 The group $H_{\Lie}^3 \left({ C^\Inf \left({ \TT^2 }\right), \RR }\right)$ 
is  at least two-dimensional. Nontrivial 3-cocycles 
can be defined as follows :
$$c_0 (f, g, h) = 
\int_{ }^{ }\!\!\int_{\TT^2 }^{ } ( \{ f, g \} h ) \om$$
and
$$
c_1 (f, g, h) = 
\Alt_{ f, g, h  }
\int_{ }^{ }\!\!\int_{\TT^2 }^{ }
\left({ { \part^3 f \over \part p^3 } 
{ \part^3 g \over \part q^3 }  -
3 { \part^3 f \over \part p^2 \part q } 
{ \part^3 g \over \part q^2 \part p }
+
3 { \part^3 f \over \part p \part q^2 }
 { \part^3 g \over \part q \part p^2 } -
{ \part^3 f \over \part q^3 } 
{ \part^3 g \over \part  p^3 }
}\right)
h \cdot \om,$$
where $f, g, h \in N$.
\end{thm}

\begin{rem}
One of the most important features  of the algebra
$ C^\Inf \left({ \TT^2 }\right)$ is the existence of an 
ad-invariant pairing (Killing form) :
$\left\langle{ f, g }\right\rangle =
\int_{ }^{ }\!\!\int_{\TT^2 }^{ }
(f \cdot g) \om$.
This allows one to describe the above cocycles in the following
 alternative way.
\end{rem}

To prove \thmref{c0c1} we introduce a map $\t$ which we
are going to exploit intensively in Section 4:
$$\theta : C^k (N, N) \ra C^{k +1} (N, \RR)$$
(where $C^k{(N, V)}$ stands for the additive group of 
$V$-valued $k$-cochains on $N$). Namely,
$\theta (c) \left({ f_1, \dots, f_{k+1} }\right) =
\Alt_{ f_1, \dots, f_{k+1} }^{ }
\left\langle{ c  \left({ f_1, \dots, f_{k} }\right) ,
f_{k+1} }\right\rangle$.
Here and below $\Alt$  denote the sum over all shuffles of arguments with
the corresponding  signs (in the case above the sum consist of $(k+1)$ 
summands due to skewsymmetry of $c$).

\begin{lem}\label{dthetalem} 
Let $c\in Z^k (N, N)$ be a closed cochain such that
\be\label{alt=0} 
\Alt_{f_1,\dots,f_{k+2}} \<\{f_1,f_2\} , c(f_3,\dots,f_{k+2})\> =0 
\end{equation}
(sum over $(k+2)(k+1)/2$ summands). Then $\theta(c)\in Z^{k +1} (N, \RR)$.
\end{lem}

\begin{pf} The exterior product of the identity cochain
$b=(\id:N\to N)\in C^1(N,N)$ with $c$ gives a cochain in 
$C^{k+1}(N,N\tens N)$
\[ (b\boxtimes c)(f_1,\dots,f_{k+1}) = 
\Alt_{f_1,\dots,f_{k+1}} f_1\tens c(f_2,\dots,f_{k+1}) .\]
Its coboundary is 
\[ d(b\boxtimes c) = db\boxtimes c - b\boxtimes dc 
= db\boxtimes c = \{,\}\boxtimes c\]
\[ (\{,\}\boxtimes c)(f_1,\dots,f_{k+2}) = \Alt_{f_1,\dots,f_{k+2}}
\{f_1,f_2\}\tens c(f_3,\dots,f_{k+2}) \]
since $db$ is the bracket $\{\,,\,\}$ itself. Applying the homomorphism
$\<\,,\,\>: N\tens N\to \R$ we get the cochain $\theta(c)$ with 
the coboundary
\be\label{d(theta)=Alt} 
d(\t(c)) (f_1,\dots,f_{k+2}) =
\Alt_{f_1,\dots,f_{k+2}} \<\{f_1,f_2\} , c(f_3,\dots,f_{k+2})\> 
\end{equation}
which vanishes by assumption.
\end{pf}

\begin{example}
Denote by $\xi_p \in H^1 (N, N)$ (resp. $\xi_q$) the following outer 
derivation of the algebra $N$: $\xi_p {(*)} = \{p, * \} : N \ra N$
(resp. $\xi_q {(*)} = \{q, * \}$). Then the 2-cocycle $c_p=\theta(\xi_p)$ 
(resp. $c_q=\theta(\xi_q)$) is 
$c_p (f, g) = \left\langle{ \xi_p (f), g }\right\rangle$ 
(resp. $c_q(f,g) = \<\xi_q(f),g\>$). The condition~\eqref{alt=0} 
follows from the skew-symmetricity of $\xi_p$.
\end{example}

\begin{example}
The wedge product $d_0 = \xi_p \wedge \xi_q =\{,\}$ is an $N$-valued 
2-cocycle. The property~\eqref{alt=0} holds by the Jacobi identity. And
 it is easy to see, that $\theta (d_0) (f, g, h) = c_0 (f, g, h)$.
Note that even though $d_0$ is a trivial 2-cocycle (i.e. is a 2-coboundary), 
its image under the homomorphism $\theta$ is a nontrivial 3-cocycle on $N$. 
The 2-cocycle $d_0$ corresponds to the deformation of $N$ evolved by a 
dilation of the area form $\om$.
\end{example}

In the Section 4 we extend this construction to the algebras of
pseudodifferential operators and to their $q$-analogues.

A common feature of these examples is the notion of invariant cocycle on a 
Lie algebra with an ad-invariant form. This notion (defined and intensively
used in \cite{[L-R2]})  generalizes naturally the idea of skew-symmetric
1-cocycle. Let $\gg$ be a Lie algebra with an ad-invariant form $\<\,,\,\>$,
one says that a cocycle $c\in C^k(\gg,\gg)$ is invariant if the map 
$\t'(c)$ defined through 
$\t'(c)(x_0,\dots,x_k) = \<c(x_0,\dots,x_{k-1}) , x_k\>$ is antisymmetric.
One checks easily that invariant cocycles form a subcomplex in
$C^*(\gg,\gg)$ (denoted by $C^*_I(\gg,\gg)$), and $\t'$ (proportional to 
$\t$) defines a morphism of complexes:
\[ \t': C_I^k(\gg,\gg) \to C^{k+1}(\gg,\R) .\]
So one has naturally the notion of invariant cohomology denoted 
$H_I^k(\gg,\gg)$ with a map still denoted by $\t'$
\[ \t': H_I^k(\gg,\gg) \to H^{k+1}(\gg,\R) .\]

Notice that the cohomology class of $d_0=[\,,\,]$ is nontrivial in
$H^2_I(\gg,\gg)$, but is trivial in $H^2(\gg,\gg)$. The same phenomenon 
occurs  for semisimple classical Lie algebras: the Lie bracket defines a
nontrivial element in $H^2_I(\gg,\gg)$ whose image in $H^3(\gg,\R)$
generates the latter group (see e.g. Section 2.2).

\begin{example}
The 3-cocycle $c_1$ is obtained by the same homomorphism $\theta'$ from a
nontrivial 2-cocycle
$d_1 \in H^2 (N, N)  ;
d_1 =
{ { \part^3 f \over \part p^3 } 
{ \part^3 g \over \part q^3 }  -
3 { \part^3 f \over \part p^2 \part q } 
{ \part^3 g \over \part q^2 \part p }
+
3 { \part^3 f \over \part p \part q^2 }
 { \part^3 g \over \part q \part^2 p } -
{ \part^3 f \over \part q^3 } 
{ \part^3 g \over \part  p^3 }}$.
This cocycle $d_1$ is responsible for a formal deformation of
 the Poisson algebra; in fact, it is the third term of the 
 Moyal deformation (called also deformation quantization) of the Poisson
algebra $N$, obtained by extending the Heisenberg law $\{p,q\}=1
\rightarrow [\hat p, \hat q]=Id$. Another notation for this cocycle is
$d_1 (f, g) = \left(\mu{ \bgt^3 }\right) (f \ot g)$
(here $ \bgt =
\left({ 
{\part \over \part p} \ot {\part \over \part q}
-{\part \over \part q} \ot {\part \over \part p}
 }\right)$ and $\mu$ denotes the multiplication).
One checks easily that this cocycle is invariant: if one uses
trigonometric representation for functions on $\TT^2$, say 
$ L_{\alpha} = - \exp \left({ i \left({ \alpha_1 q + \alpha_2 p }\right)
  }\right)$, $\alpha\in\Z^2$, then
\[ \int_{\TT^2} d_1(L_\alpha,L_\beta) L_\gamma \omega =
\d_{\alpha+\beta+\gamma,0} (\alpha\times\beta)^3 ,\]
where $\alpha\times\beta = \alpha_2\beta_1 - \alpha_1\beta_2$.
Invariance is then immediate. Linear independence of the classes of
$c_0$ and $c_1$ in $H^3(N,\R)$ can be also checked through this formula.
Nontriviality of these 2-cocycles is equivalent to nontriviality 
of the corresponding Lie algebra deformation (see, e.g. \cite{[Fu]}).
\end{example}

\begin{rem}
The cohomology class of $d_1$ in $H^2 (N,N)$ is sometimes called the 
Vey class after Vey's fundamental contribution to the subject \cite{[Ve]}.
It admits a global analogue 
for any symplectic manifold, and is often denoted by $S^3_\Gamma$. 
This Moyal deformation is linked with symbols of differential operators via 
the Weyl ordering.
For arbitrary surfaces the invariance
of that cocycle is conjectured, but to the best of our
knowledge has not been proved yet.
\end{rem}

I. Gelfand and O. Mathieu \cite{[G-M]} found a nice construction of higher 
cocycles of the Poisson algebra $N$. We generalize them to the algebras
of pseudodifferential or $q$-pseudodifferential operators and 
sine-algebras discussed below.
This approach can be explained through the following approximation scheme.

\subsection{Approximation of $N = C^\Inf \left({ \TT^2 }\right)$ by 
$gl(m)$ as $m \ra \Inf$} (following \cite{[H-O-T]}). 

Let us fix some odd $m$ and consider the following two unimodular 
matrices in $gl(m)$:
\[ E \equiv
\begin{pmatrix}
1& & & &\\
&\ve & & 0 &\\
& & & &\\
& 0 & &\ddots &\\
& & & &\ve^{m-1}
\end{pmatrix} 
\qquad \text{and} \qquad H \equiv
\begin{pmatrix}
0& 1& & &\\
& & & 0&\\
& & \ddots& \ddots&\\
&0 & & &1\\
1& & & &0
\end{pmatrix}
\]
where $\ve$ is a primitive $m^{{\mathop{\rm  th}\nolimits} }$ root of unity
(say, $\exp\left({ { 4 \pi i \over m} }\right)$). They obey the identities 
$HE = \ve EH, E^m = H^m = \un$. Then the unitary unimodular matrices
$$\Ja_{ \left({k_1, k_2 }\right) } =
\ve^{k_1 \cdot k_2 /2} E^{k_1} H^{k_2}$$
 span the algebra of $gl(m)$. Note that
$\Ja_{\left({k_1, k_2 }\right) }^{-1} = \Ja_{\left({- k_1, - k_2 }\right) }$
and $\tr \Ja_{ \left({k_1, k_2 }\right) } = 0$ except for
$k_1 = k_2 = 0\ (\mod m)$. 
The set of $\Ja$'s is closed under multiplication:
$$\Ja_k \Ja_\ell = \ve^{k \ts \ell /2} \Ja_{k + \ell}
\qquad
\left({ 
k = \left({ k_1, k_2 }\right) ,
\ell = \left({ \ell_1, \ell_2 }\right),
k \ts \ell =
k_2 \ell_1 - k_1 \ell_2
 }\right).$$
Therefore 
$\{\Ja_k \} $ satisfy the commutation relations
\be
\left\lbk{ \Ja_k, \Ja_\ell  }\right\rbk
= 2 i \sin 
\left({ { 2 \pi ( k \ts \ell )\over m } }\right)
\Ja_{k + \ell }
\label{(1)}
\end{equation}
The real subalgebra $\frak a(m)$ of $gl(m)$ spanned by $i\Ja_k$ is 
isomorphic to $u((m+1)/2,(m-1)/2)$, the Lie algebra of the group of 
matrices unitary with respect to hermitian metric in $\C^m$ with $(m+1)/2$ 
positive and $(m-1)/2$ negative squares. Indeed, 
\[ (-1/m) \Tr (i\Ja_k\cdot i\Ja_l) = \delta_{k+l,0} ,\]
which implies that the signature of the Killing form on $\frak a(m)/\R\Ja_0$
is 0. For $su(k,l)$ this signature is $(k-l)^2-1$.

As $m \ra \Inf$ this algebra goes to the algebra
$\left\lbk{ L_k, L_\ell }\right\rbk =
(k \ts \ell ) L_{k + \ell}$ through the identification 
$(m / 4 \pi i) \Ja_k \mapsto L_k$. The latter object is exactly 
 the Poisson algebra of Hamiltonian functions 
on $\TT^2$:
$$
L_{\left({ k_1, k_2 }\right)} =
- \exp 
\left({ i
\left({ k_1 q + k_2 p }\right)
  }\right),$$
which is modulo constants the same as
the algebra of the corresponding Hamiltonian vector fields:
$$L_{\left({ k_1, k_2 }\right)} =
 i \exp 
\left({ i
\left({ k_1 q + k_2 p }\right)
  }\right)
\left({ k_1 { \part \over \part p} -
k_2 { \part \over \part q}
}\right).$$
The algebra \eqref{(1)} is a non-extended case (``cyclotomic family'') 
of an infinite dimensional ``sine-algebra'' \cite{[F-F-Z1],[F-F-Z2],[F-Z]}:
\be
\left\lbk{ \Ja_k, \Ja_\ell }\right\rbk =
r \sin (2\pi (k \ts \ell)/\l) \Ja_{k + \ell} +
( a \cdot k) \delta_{k + \ell, 0}
\label{(2)}
\end{equation}
where the constant $\l$ is not necessarily integer, but is an arbitrary
complex number,
$ a = \left({ a_1, a_2 }\right)$ is a fixed plane vector, $r\in\CC$,
$k  = \left({ k_1, k_2 }\right),
\ell = \left({ \ell_1, \ell_2 }\right)$ are not integers modulo $m$, 
but belong to $\ZZ^2$, and $(a\cdot k) = a_1k_1+a_2k_2$.

Unlike the (simple) $sl (m)$ algebra, the (infinite dimensional) 
sine-algebra and the 
Poisson algebra $\{ L_k\} $ have two nontrivial central extensions.
 Thus 2-cocycles on 
the sine-algebra with an arbitrary $\l$ disappear when we truncate it to the
finite-dimensional object. However, the situation with real- (or complex-)
valued 3-cocycles is different. It is well-known that the algebra 
$su((m+1)/2,(m-1)/2)$ as any simple Lie algebra has a 
1-dimensional group $H^3 (su((m+1)/2,(m-1)/2), \RR)$.

\begin{prop}
 The Lie 3-cocycle on $u((m+1)/2,(m-1)/2)$  generating 
$H^3_{Lie} (u((m+1)/2,(m-1)/2)) , \RR)$ goes to the 3-cocycle $c_0$ 
(up to a multiplicative constant) of \thmref{c0c1} as $m \ra \Inf$.
\end{prop}

\begin{pf} The cocycle on $\frak a(m)$ is given by the completely
antisymmetric expression $\mu (A, B, C) =a\Tr ([A, B] C)$, 
$a=(4\pi i)^2 m^{-3}$. Note, that due to a nondegenerate pairing on 
$\frak a(m)$: $\left\langle{ A, B }\right\rangle = \Tr (AB) $ 
the cocycle $\mu$ can be obtained from an invariant 2-cocycle 
$\eta \in Z^2 (\frak a(m), \frak a(m))$: $\eta (A, B) = [A, B]$ via 
$\t(\eta)=\mu$ (see Section 2.1). As $m \ra \Inf$ the commutator 
of matrices $[ \ , \ ]$ goes to the Poisson bracket, and 
$a\Tr ( \ \cdot \ )$ goes to  the ad-invariant trace on $N$ defined by 
$\Tr (f)= \int_{\TT^2 }^{ } f \om$.
Indeed, 
$\left\lbk{ L_f, L_g }\right\rbk = L_{ \left\lbc{ f, g }\right\rbc }$, 
and in the limit the trace becomes the zero Fourier mode of a Hamiltonian
function. 
\end{pf}

The coordinate expressions of the cocycles are 
$$\mu 
\left({ \Ja_\a, \Ja_\b, \Ja_\g }\right) = \frac{m}{2\pi} \sin\left(
\frac{2\pi(\a  \ts \b)}m \right)\, \delta_{\a + \b + \g,0}$$
($ \a, \b, \g \text{ are } \mod m$), and
$$c_0 
\left({ L_\a, L_\b, L_\g }\right) =
(\a  \ts \b)\delta_{\a + \b + \g,0}$$

\section{Adjacency diagram of the algebras 
$C^{\Inf} ( \TT^2 ), \protect\sin_{\l}, \psi DO, \psi DO_q$} 
In the preceding section we showed that the 3-cocycle on 
$u((m+1)/2,(m-1)/2)$ survives under the limit $m \ra \Inf$. In the next 
section we will see how it deformes when
one quantizes the Poisson algebra $N$ into $\psi DO$ or $\psi DO_q$.

The diagram of various algebras discussed in this paper is as follows
\[
\begin{array}{rccccccc}
u((m+1)/2,(m-1)/2) 
& \lhook\joinrel\mathrel{\hfl{}{}}
& \sin_\l 
& 
\hookrightarrow &\psi DO_q \\
{\scriptstyle m \ra \Inf}\nquad & \searrow & \vfl{}{\l \ra \Inf} &  &
\vfl{}{q \ra 1} \\
 & &C^\Inf \left({ \TT^2 }\right) & 
\hfll{}{ \{ \ , \ \} \la [ \ , \ ] } &  \psi DO 
\end{array}
\]
In this section we  define all the objects and meaning of all arrows 
in this diagram. We recall also certain well- and less-known facts 
about the objects involved.

\subsection{Algebras $u((m+1)/2,(m-1)/2)$ and $\protect\sin_\l$}
In the last section it was mentioned that the limit $ m \ra \Inf$
of the structural constants of $u((m+1)/2,(m-1)/2)$,
is to be considered in the framework of generic infinite dimensional
sine-algebras $\sin_\l$ (with rational or irrational $\l$):
$$
\left\lbk{ \Ja_k, \Ja_\ell }\right\rbk =
{\l \over 2 \pi}
\sin \left({ {{2 \pi (k \ts \ell)}\over \l} }\right)
 \Ja_{k + \ell}.$$

\subsection{Algebra of pseudodifferential operators on the line 
and on the circle.} 

The algebra of $\psi DO$ is a quantization of the algebra 
$N = C^\Inf \left({ \TT^2 }\right)$ where $q$ is replaced by $x$ and $p$ is
replaced by
${\part  \over  \part x}$. More formally,

\begin{defn}
A ring $\psi DO$ of pseudodifferential symbols is the ring of formal
series 
$A (x, \part) = 
\sum_{ - \Inf }^{ n} a_i (x) \part^i$ with respect to $\part$, where 
$a_i (x)  \in \CC \left\lbk{ x, x^{- 1} }\right\rbk$ (or 
$C^{\infty}{ (S^1, \RR \text{ or } \CC )}$), and the variable $\part$ 
corresponds to $d / dx$.
The multiplication law in 
$\psi DO$ is given by the commutation relations
$$\part \circ f(x) = f(x) \part + f' (x)$$
$$\part^{-1} \circ f(x) = f(x) \part^{- 1} -
f' (x) \part^{- 2} + f'' (x) \part^{- 3} - \cdots$$
\end{defn}

These relations define the usual composition law on the subalgebra
${DO} \sbs \psi DO $ of {\it dif\-fe\-rential} operators (i.e. 
on polynomials with respect to $\part$) and they can be unified in one as
\be
\part^s \circ u(x) = 
\sum_{ \ell \geq 0}^{ }
\left({ { s\atop \ell} }\right)
u^{(\ell)} (x) \part^{s - \ell},
\label{(3)}
\end{equation}
where $\left({ { s\atop \ell} }\right) =
{s(s - 1) \cdots (s - \ell + 1) \over \ell ! }$. The product 
determines the natural Lie algebra structure on 
$\psi DO$: $[A, B] = A \circ B - B \circ A$. The space $\psi DO$ with
this Lie algebra structure will be denoted by $\G$.

Moreover, there are  operators res: 
$\psi DO \ra \CC \left\lbk{ x, x^{- 1} }\right\rbk$ 
(or $C^\Inf \left({ S^1 }\right)$) and Tr:$\psi DO\ra \CC$
on the ring 
$\psi DO $:  res $
\left({ \sum a_i D^i }\right) = a_{- 1} (x)$, and Tr$A=\int$ res $A$.
 The main property of Tr is
Tr $[A, B] = 0$ for arbitrary 
$A, B \in \psi DO$ (here and below we integrate over the circle
$S^1$).

\begin{rem}
The multiplication formula above is defined not only for integer values of
$s$, but for fractional and complex values as well. 
One can define an associative product on $\psi DO$ depending on a 
parameter $h$. The Leibnitz rule for multiplication of symbols in 
$\psi DO$ gives the following general formula equivalent 
 for $h = 1$ to the rule~\eqref{(3)} above:
$$A(x, \xi) \build {\circ }_{ h}^{ } B (x, \xi) =
\sum_{ n \geq 0 }^{ }
{h^n  \over n! } A_\xi^{(n)} (x, \xi) B_x^{(n)} (x, \xi)$$
where $A_\xi^{(n)} = {d^n  \over d \xi^n } A (x, \xi),$ 
$B_x^{(n)} = {d^n  \over d x^n } B(x, \xi)$. 
Let $\Phi_t:\G \to \G $ defined through 
$\Phi_t(a(x)\xi^p) = a(x) t^{p-1} \xi^p$ be a family of bijective maps
for $t\in ]0,1]$, singular for $t=0$. Set 
$[A,B]_t = \Phi_t^{-1} [\Phi_t(A), \Phi_t(B)]$. One has 
$[A,B]_t =\{A,B\}+t\dots$. So when $t\to0$ $\lim[A,B]_t = \{A,B\}$.
One says that the commutator of pseudodifferential symbols contracts
onto the Poisson bracket.
\end{rem}

\subsection{The algebra of $q$-pseudodifferential symbols.}
\subsubsection{Associative algebra structure.}
The notion of the $q$-analog of the derivative (and any differential) 
operator (see, e.g., \cite{[Ka]}) is extended here to the case of 
pseudodifferential operators. We follow the framework of the previous 
subsections stressing the difference between the classical and 
$q$-deformed case.

Let $F$ be the algebra $\CC \left\lbk{ x, x^{- 1} }\right\rbk$ and 
$q \in \CC^\ts$. It will be the main algebra of coefficients for us, 
the reader can easily reproduce the results for
the algebra 
$F' = C^\Inf \left({ S^1 }\right) $ in the case
$q \in \CC, | q | = 1$ etc.
We use the following notations for $q$-numbers
$$(n)_q = 
{q^n - 1 \over q - 1}$$
$$
\left({ {m \atop \ell} }\right)_q =
{(m)_q (m - 1)_q \cdots (m - \ell + 1)_q
 \over (1)_q (2)_q \cdots (\ell)_q} .$$
Define the $q$-analog of the derivative by the following
\[ D_q f(x) = {f(qx)-f(x) \over q-1} \]
for $f\in F$. It will be useful to define also the shift
\[ \tau f(x) = f(qx) , \qquad \tau^\beta f(x) = f(q^\beta x) .\]
Clearly, $\tau$ commutes with $D_q$.
So, $D_q$ is a $q$-derivative in the following sense:
$$D_q (fg) = D_q (f) g + \tau (f) D_q (g).$$

\begin{defn}
An algebra $\psi DO_q$ of $q$-pseudodifferential operators is a
vector space of formal series
$$\psi DO_q =
\left\lbc{  A \left({ x, D_q }\right) =
\sum_{- \Inf  }^{ n} u_i (x) D_q^i \bigm| u_i \in F }\right\rbc$$
with respect to $D_q$. The multiplication law in 
$\psi DO_q$ is defined by the following rule: $F$ 
is a subalgebra of $\psi DO_q$ and there are commutation relations:
$$D_q \circ u =
\left({ D_q u}\right)
+ \tau (u) D_q, 
\qquad u \in F
$$
\be
D_q^{- 1} \circ u =
\sum_{ k \geq 0 }^{ }
( - 1)^k 
\left({ \tau^{- k - 1} 
\left({ D_q^k u }\right)
 }\right) D_q^{- k -1},
\qquad u \in F, F'.
\label{(4)}
\end{equation}
\end{defn}

Each term of the product of two Laurent series in $D_q$ is found by applying
these rules finite number of times. The formula~\eqref{(4)} is built so that
$D_q^{-1} \circ D_q \circ u = D_q \circ D_q^{-1} \circ u = u$.
For $q = 1$ these formulas recover the ``classical'' definition of
multiplication law in the algebra of pseudodifferential operators 
$\psi DO$.

The commutation rule for $D_q^n $ (with any integer $n$) and $u(x)$ join
these formulae in one :
\be
D_q^n \circ u =
\sum_{\ell \geq 0 }^{ }
 \left({ { n \atop \ell} }\right)
\left({ \tau^{n - \ell} 
\left({ D_q^\ell u }\right)
 }\right) D_q^{n - \ell} .
\label{(5)}
\end{equation}

\begin{prop}
The $q$-analog of the Leibnitz rule of multiplication of two
$q$-pseudo\-dif\-ferential operators 
$A \left({ x, D_q }\right), B \left({ x, D_q }\right)$ can be written
as the following operation on their symbols
\be
A \left({ x, D_q }\right) \circ 
B \left({ x, D_q }\right) =
\sum_{ k \geq 0 }^{ }
{1 \over (k) ! } 
\left({ { d^k \over d D_q^k}  A}\right) *
\left({ D_q^k B }\right)
\label{(6)}
\end{equation}
where
${d^k \over d \xi^k} A({x, \xi})$ is the $q$-derivative of $A$ with 
respect to the second argument (for $A = f(x) D_q^\a$ it is 
${ d^k \over d D_q^k}  A =
{ d^k \over d D_q^k}  \left({f D_q^\a }\right) =
{ (\a) !\over (\a - k) !} f D_q^{\a - k}$), and
$D_q^k B \left({ x, D_q }\right)$ is the $q$-derivative of $B$
 defined above, and finally, the multiplication $*$ of symbols is
defined with the following commutation rule of the generators:
$$f * D_q = f D_q, \qquad D_q * f = \tau (f) D_q,
\qquad D_q^{-1} * f = \tau^{-1} (f) D_q^{- 1}.$$
\end{prop}

\begin{pf}
Straightforward verification of the formula~\eqref{(6)} for the product
$D_q^n \circ u (x) $ gives the same answer as \eqref{(5)}.
\end{pf}

\subsubsection{Non-commutative residue for $\psi DO_q$.} \label{Traces}
1. The operation $res:{\psi DO}_q \ra F$ is defined by
$$\res \left({  
\sum_{i = - \Inf }^{ n }  u_i (x) D^i_q
}\right) = u_{- 1} (x)$$
generalizing definition for classical pseudodifferential symbols \cite{[Wo]}.

\medskip
2. The integral along the circle $S^1$ gives a linear functional
$\int : F \ra \CC$, $\int x^n=\d_{n,0}$, on the function algebra $F$  
satisfying $\int D_q f = 0$ and $\int \tau (f) = \int f$ for all $f \in F $. 
We have also an ``integration by parts'' formula 
$\int f \tau^{-1}(D_q(g)) = - \int D_q(f) g$.

\subsubsection{Inner product on $\psi DO_q$.}
Define in $\psi DO_q$ the element $T=(q-1)D_q +1$ and denote its inverse
\[E \equiv T^{-1} = \sum_{i=1}^\infty \frac{(-1)^{i-1}}{(q-1)^i} D_q^{-i} 
= \frac1{(q-1)D_q+1}.\]

\begin{lem}\label{tau=AdT} 
The automorphism $\tau :\psi DO_q\to\psi DO_q $, 
$fD_q^p\mapsto \tau(f)D_q^p$ is inner. 
Precisely, $\tau(A) = TAT^{-1} = E^{-1}AE$.
\end{lem}

\begin{pf} 
Indeed, for any $f\in F$
\begin{align*}
TfT^{-1} &= ((q-1)D_q + 1)fT^{-1} \\
&= \{(q-1)(\tau(f)D_q + D_q(f)) + f\} T^{-1} \\
&= \{\tau(f) ((q-1)D_q + 1) + (q-1)D_q(f) - \tau(f) + f\} T^{-1} \\
&= \tau(f) TT^{-1} = \tau(f)
\end{align*}
and $T$ commutes with $D_q$.
\end{pf}

Define the Lie algebra ${{\tenfm G}}_q$ as the set $\psi DO_q$ of all 
$q$-pseudodifferential symbols equipped with the commutator bracket
$[A, B] = A \circ B - B \circ A$.

\begin{thm}\label{res(ABE)} 
 The bilinear form 
$ \left\langle{ A, B }\right\rangle = \int \res (ABE)$ is an ad-invariant 
symmetric non-degenerate form on the Lie algebra ${{\tenfm G}}_q$.
\end{thm}

\nn 
To prove it consider the bilinear form 
$(A, B) \mps \int \res (A \circ B)$ on the algebra 
${{\psi DO}}_q$. It is not symmetric but it has the following symmetry
property.

\begin{prop}
 For any $A, B \in \psi DO_q$
$$\int \res (A \circ B) =  \int \res (B \circ \tau (A)),$$
where the map $\tau : {\psi DO}_q \ra {{\psi DO}}_q$ 
is the following automorphism of the algebra
${\psi DO}_q$:
$$\tau \left({ f D_q^b }\right) = \tau(f) D_q^b $$
\end{prop}

\begin{pf} 
Explicitly the form $\int\res$ is given by
\begin{align*}
\int \res \left({ f D_q^a \circ g D_q^b }\right) &= \int\res 
\bigl(f\sum_{k\ge0} \binom ak \tau^{a-k}(D_q^k g) D_q^{a-k+b} \bigr) \\
&= \left({ {a \atop a + b + 1} }\right) \int f \tau^{- b - 1}
\left({ D_q^{a + b + 1}  g}\right)
\end{align*}
for $a + b + 1 \geq 0$, and it  vanishes otherwise. On the other hand assuming
$a+b+1\ge0$ we find  also that
\begin{align*}
\int \res \left({ \tau^{- 1} (g) D_q^b \circ f D_q^a}\right) &=
\binom b{a+b+1} \int \tau^{-1}(g) \tau^{-a-1}(D_q^{a+b+1}f) \\
&= \frac{b(b-1)\dots(-a)}{(a+b+1)!} 
\int \tau^{-b-1}(g) \tau^{-a-b-1}(D_q^{a+b+1}f) \\
&= \frac{(-b)(-b+1)\dots a}{(a+b+1)!} 
\int D_q^{a+b+1}( \tau^{-b-1}g) f 
\end{align*}
which coincides with the expression above.
\end{pf}

Now \thmref{res(ABE)} immediately follows from  \lemref{tau=AdT}.

\subsubsection{Twisted loop algebra.}\label{KaRa} 
The algebra  $\psi DO_q$ can also be described as a twisted loop algebra 
following \cite{[K-R]}. This construction, in short, is as follows.

For an associative algebra $A$ and its automorphism $\s$ one defines the
 twisted loop algebra $A_\s [x,x^{-1}]$. As a vector space this is the 
space of Laurent polynomials $\sum_{j\in \ZZ} x^j a_j$, $a_j \in F$. The 
associative product is defined by the rule:
$$(x^ma)(x^nb)=x^{m+n}\s^n(a)b.$$

Of course, when $\s$ is the identity automorphism, one recovers the usual 
notion of the loop algebra. (Those algebras are purely algebraic versions
of the well-known cross-product in functional analysis.)

With any trace $Tr$ on $A$ one can associate a central extension of
$A_\s [x,x^{-1}]$ defined by the following 2-cocycle:
$$\psi_{\s , Tr}(x^ma,x^nb) = -\psi_{\s , Tr}(x^nb,x^ma)  = 
Tr((1+\s +\dots+\s^{m-1})(\s^{-m}(a)b))\,\d_{m+n,0}$$
if $m$ is positive, and 0 if $m=0$.

In order to obtain $q$-pseudodifferential symbols simply take
$A=\CC [\xi, \xi^{-1}]]$ and for some $q\in\CC^*$ set 
$\s_q(\xi)=q\xi+1 $ so that $\s_q^n(\xi)=q^n\xi+(n)_q$.

We get then an isomorphism of associative algebras
$$I: A_\s [x,x^{-1}] \ra \psi DO_q$$
defined on generators as follows:
$$I(x^n)=x^n \ \text{ and } \ I(\xi)=D_q .$$

In the particular case of $q=1$, one recovers the associative algebra of
pseudodifferential operators by assigning to $\xi$ the operator
$f\mapsto x\cdot {\part f}/{\part x}$.

\subsection{The sine-algebra and $\psi DO_q$.}
Here we identify $q$-analogs of $\psi DO$ and sine-algebra, following  
M. Golenishcheva-Kutuzova (see \cite{[G-L]}), and show that the invariant 
trace  on $\psi DO_q$ we have constructed above becomes the canonical 
invariant trace on the sine-algebra. We are grateful to 
M. Golenishcheva-Kutuzova for explaining to us this alternative approach
to the trace description (\cite{[Gol]}).

To define an associative product on elements of the sine-algebra, consider
the ``quantum torus'', i.e. the $C^*$-algebra $A_h$ generated by two unitary 
operators $U_1$ and $U_2$, satisfying the relation 
$U_2 U_1 = q U_1 U_2$ $ (q = e^{ih})$.

An arbitrary element of $A_h$ can be written in the form of a formal series
$f = \sum_{ n_1, n_2 \in \ZZ }^{ } 
f_{(n_1, n_2)} U_1^{n_1} U_2^{n_2}$. The Lie algebra structure on $A_h$ is
defined by 
$$[f, g] = f * g - g * f =
\sum_{n, m \in \ZZ^2 }^{ } 
f_{(n_1, n_2)} g_{(m_1, m_2)}
\left({ q^{m_1 n_2}  - q^{m_2 n_1} }\right)
U_1^{n_1 + m_1} U_2^{n_2 + m_2}.$$

One can construct a two-dimensional central extension of this Lie algebra
following the prescription of Section 2.1 above:
the sine-algebra $A_h$ admits an ad-invariant trace 
$\eta \left({ \sum_{ n_1, n_2 }^{ } 
f_{n_1, n_2} U_1^{n_1} U_2^{n_2} }\right) = f_{0,0} $. 
This trace allows one to associate a 2-cocycle with scalar
values to  every outer derivation of the algebra.
 
Let $a=(a_1,a_2)\in \CC$ and $L_a$ be a derivation defined by
$L_{(a_1, a_2)}(U_1) = a_1 U_1$,  $L_{(a_1, a_2)}(U_2) = a_2 U_2$.
Then the corresponding 2-cocycle
is nothing but $c(f, g) = \eta \left({ L_a f * g  }\right)$.
Having chosen the basis in $\hh A_h = A_h \op \CC c $ in the form 
$\Ja_{k_1 k_2} = q^{k_1 \cdot k_2 /2} U_1^{k_1} U_2^{k_2}$ 
we come to the commutation relations of the sine-algebra~\eqref{(2)}
with $q=e^{4\pi i/\l}$, see \cite{[G-L]}.

Consider now the homomorphism $\phi:\sin_\l \hookrightarrow \G _q$
defined through $\phi(U_1)=x$, $\phi(U_2)=D_q + {1 \over q-1}$.

\begin{prop}
The $\phi$-pullback of the ad-invariant trace on $\G _q$ induces  the
ad-invariant trace $\eta$ on the sine-algebra $\sin_\l$ 
(with $q=e^{4\pi i/\l}$) up to a constant multiple.
\end{prop}

\begin{pf} Indeed,
\begin{align*}
\Tr\phi(U_1^kU_2^l) &= \Tr\Bigl(x^k\bigl(D_q+{1\over q-1}\bigr)^l\Bigr) \\
&= \int\res \Bigl(x^k \bigl(D_q+{1\over q-1}\bigr)^l
\frac1{(q-1)D_q+1)}\Bigr) \\
&= {1\over q-1} \int\res\Bigl(x^k \bigl(D_q+{1\over q-1}\bigr)^{l-1}\Bigr)\\
&= {\d_{l,0}\over q-1} \int x^k = {1\over q-1} \d_{l,0} \d_{k,0}
\end{align*}
\end{pf}

\begin{rem}
The invariant inner product on both algebras is defined by the trace via
$$ \left\langle{ A, B }\right\rangle = \Tr (A  B).$$
\end{rem}

\subsection{Bialgebra structure}
We recall that the classical algebra $ {\got G}$ of
pseudodifferential symbols can be equipped with a Lie bialgebra structure.
Indeed, the algebra $ {\got G}$ as a linear space is a direct sum of two
natural subalgebras: $ {\got G}_+$ consisting of differential
operators $\sum_{j \ge 0}^{} a_j D^{j}$, and $ {\got G}_- =  {\got
G}_{\rm Int}$ consisting of integral symbols $\sum_{j=-1}^{- \infty} a_j
D^{j}$.

The triple of algebras $({ {\got G}},{
{\got G}}_+, { {\got G}}_- )$ is a {\it Manin triple }(or, equivalently,
$\widetilde { {\got G}}_{\rm Int} = \widetilde { {\got G}}_-$ is a 
{\it Lie
bialgebra)} (see, e.g. \cite{[Dr]}).  This means that 1) the algebra
$ {\got
G}$ has an ad-invariant nondegenerate inner product (``Killing form"):
$$ \big \langle A, B \big \rangle =
 \int \res (A \circ B) $$
for $A,B \in  {\got G}$, 2) 
as a linear space
$ { {\got G}} =  { {\got G}}_+ \oplus  { {\got G}}_-$, and 3) both the
subalgebras are isotropic with respect to this Killing form.
The Lie group $ G_{\rm Int}$ corresponding to the Lie bialgebra 
$\what { {\got G}}_{\rm Int} = \widetilde
{ {\got G}}_-$ has a natural Lie-Poisson structure.

The Lie algebra 
$s{\got G}_q =\{ A= \sum_{k=-\infty}^{n} u_k (x) T^k \mid \Tr A=0 \}$, 
where $T=(q-1)D_q+1$, is a sum of two isotropic subalgebras: 
of $q$-differential
( $\G^+_q = \{ \sum_{k=0}^n u_k (x) T^k \mid u_0\in x\C[x]  \} $ ) 
and of $q$-integral operators 
( $\G^-_q = 
\{\sum_{k=-\infty}^{0} u_k (x) T^k \mid u_0\in x^{-1}\C[x^{-1}]\}$ )
which are dual to each other  with respect to the invariant 
inner product on $s{\G}_q$. Indeed, the commutation relation $Tx=qxT$
implies that $(T^mx^n)_{m,n}$ and $(x^{-n}T^{-m})_{m,n}$ are two bases
dual to each other with respect to this form. Therefore,
$(s\G_q,\G^+_q,\G^-_q)$ is also a Manin triple and $s\G_q$ is a Lie double.

\begin{rem}
 A likewise Manin triple there exists for the extension
$\widetilde { {\got G}}$ of ${ {\got G}}$ by a central and the 
corresponding cocentral terms (see Section 4). The corresponding 
Poisson structures on the nonextended and extended Lie groups
$ G_{\rm Int}$ are respectively the Benney and quadratic
Gelfand-Dickey structures \cite{[K-Z]}.
\end{rem}

\subsection{Quantum and classical $W_\infty$-algebras}

The Lie algebras of $W_\infty$-type (having ge\-nerators of all spins up to
infinity) appeared recently in conformal field theory as the large $n$
limit of Zamolodchikov's quadratic $W_n$-algebras.
Deep relations observed between these objects often look very natural and 
transparent after translation of them into the language of symbols.
A ``short dictionary'' for some of these objects is just like this.

The (quantum) $W_{1+\infty}$ algebra is isomorphic to the
algebra $\widehat {{\tenfm G}}_{DO}$
 which is the central extension of the algebra ${{\tenfm G}}_{DO}$ of
differential operators on the circle \cite{[P-R-S]}.

The quantum $W_{\infty}$ algebra is the truncated 
$\widehat {{\tenfm G}}_{DO}$ where zero
order differential operators (i.e. functions) are discarded. 
Historically this algebra appeared to be spanned by ``compositions'' of the
Virasoro (i.e. first order) generators. In this construction the
space of functions does not appear. This explains the funny from
mathematical viewpoint notation $W_{1+\infty}$ for extension of
$W_{\infty}$ by functions.

Classical $w_{\infty}$ algebra can be obtained from $W_{\infty}$
by replacement of all commutators by Poisson brackets. In the
language above this is nothing but the Poisson algebra of symbols of
differential operators.

The symbols of
differential operators are functions on a cylinder, the cotangent bundle
$T^* S^1$ of the circle. These symbols are polynomial along the
fiber $p$. Generalization of $DO$ to $\psi DO$ invites one to consider
symbols being Laurent series in $p$ with coefficients periodic in $q$
(and so defined on $\{T^* S^1 \setminus S^1 \}$). Introducing 
an exponential coordinate along the fiber $p=e^{i\cdot s}$ we come
to a complex two-torus with coordinates $s$ and $q$. In these
coordinates the outer derivation $\{\log p, -\}$ becomes 
$\{s, -\}$, the outer derivation of the Poisson algebra $N$.
Moreover, the coordinates $s$ and $q$ are symmetric in 
this representation.

Any symbol of a
differential operator is a (Hamiltonian) function on $T^* S^1$
and defines a Hamiltonian vector field on this manifold.
In such a way the group and the algebra of area preserving 
(or symplectic) diffeomorphisms on $T^* S^1$ appears in the context of
$w_{\infty}$. Moreover, any diffeomorphism of $S^1$ can be lifted to
a symplectomorphism of $T^* S^1$. In physical terms 
this means that the group $\Diff (S^1)$ is a symmetry of
$W_\infty$ algebra (see e.g. \cite{[Ju]}).

Finally, note that in the same way one can treat algebras
of (pseudo)differential operators on a higher dimensional
compact manifold $M^n$. The algebra of Hamiltonian vector
fields on $T^* M^n$ appears as its classical limit, and the
group $\Diff (M)$ as its symmetry (cf. \cite{[M-M-R]}).

\section{Lie algebra cocycles on $\G $ and $\G _q$}

\subsection{Two-cocycles.}

Recall that the commutation rule
$$
\left\lbk{ \part^\a, u(x) }\right\rbk_{}^{}
= \sum_{ k \geq 1 }^{ }
\left({ {\a  \atop k } }\right) 
u^{(k)} (x) \part^{\a - k}$$
makes sense not only for a positive integer $\a$, but also for any 
complex values of $\a$. Differentiating this identity in $\a$ at 
$\a = 0$ and using undoubtful
$$
\left.{ {d \over d \a }  }\right\vert_{\a = 0} \part^\a =
\log \part \cdot \left.{ \part^\a }\right\vert_{\a = 0}
= \log \part $$
we get commutation relation for a formal symbol $\log \part$ and a function 
$u(x) $
$$\left\lbk{ \log \part, u (x)  }\right\rbk
= \sum_{ k \geq 1 }^{ }
{(- 1)^{k+1} \over k } 
u^{(k)} \part^{- k}$$
(or, more generally 
$\left\lbk{ \log \part, u (x) \part^n }\right\rbk
= \sum_{ k \geq 1 }^{ }
{(- 1)^{k+1} \over k } 
u^{(k)} \part^{n - k}$).
Note that the result of the commutation lies in $\psi DO$. In such a way 
the symbol $\log \part$ defined an outer derivation $\xi_{\log \part}$
of the ring $\psi DO$,
$\xi_{\log \part} \in H^1 (\G , \G )$. Pairing the result of
derivation with another symbol we get a complex-valued 2-cocycle
$c \in H^2 (\G , \CC)$.

\begin{thm}[\cite{[K-K]}]
The following $2$-cocycle
\be\label{logcoc}
c(A, B) =
\int_{ }^{ } \res ([\log \part, A] \circ B) =
\int \res 
\left({ 
\sum_{ k \geq 1 }^{ }
{(- 1)^{k+1} \over k} A_x^{(k)} \part^{-k} \circ B
 }\right),
\end{equation}
gives a nontrivial central extension 
$\hat{\tenfm G}$ of the Lie algebra ${{\tenfm G}}$.
(Here $A$ and $B$ are arbitrary pseudodifferential symbols on $S^1$.)
The restriction of this cocycle to ${{\tenfm G}}_{DO}$ gives the nontrivial
central extension of ${{\tenfm G}}_{DO}$.
\end{thm}

The restriction of this cocycle to the subalgebra of vector fields 
(considered as differential operators of the first order) is the 
Gelfand-Fuchs cocycle of the Virasoro algebra. This observation 
actually implies nontriviality of the cocycle on ${{\tenfm G}}$ and 
${{\tenfm G}}_{DO}$. The ``logarithmic'' cocycle on the Lie algebra
${{\tenfm G}}_{DO}$ can be also identified with the central charge of the 
$W_{1 + \Inf}$ algebra in conformal field theory (\cite{[B-K-K]}).

\begin{rem}
Assume for a moment that $\log \part$ were an element of the algebra
${{\tenfm G}}$. Then we could define not only the commutator 
$[\log \part, A]$ but also a product $\log \part \circ A$, and rewrite the
cocycle $c(A, B) = \int \res ([\log \part, A] \circ B)$ as 
$c(A, B) = \int \res (\log \part \circ [A, B])$. The last form 
means that the cocycle  
$c(A, B)$ is a 2-coboundary (and hence is trivial) because it is a 
linear function of the commutator: $c(A, B) = 
\left\langle{ \log \part , [A, B] }\right\rangle$. Recalling 
$\log \part \notin {{\tenfm G}}$, we get a heuristic proof of  
non-triviality of the cocycle. 
\end{rem}


\begin{rem}
Notice also that the Lie algebra
${{\tenfm G}}_{DO}$ of differential operators on the circle has exactly
one  central extension (\cite{[Fe],[Li],[Wo]}), while the Lie algebra of 
pseudodifferential operators has exactly two independent central extensions 
(personal communication by B.~L. Feigin). The formula for the second cocycle 
can be written in a similar to \eqref{logcoc} form  (see \cite{[K-Z]}):
$$c' (A, B) = \int \res ([\log x, A] \circ B)$$
for coefficients $a(x) \in \CC \ [x, x^{-1}]$ or
$$c' (A, B) = \int \res ([x, A] \circ B)$$
for coefficients 
$a(x) \in C^\Inf (S^1)$.

Here $x$ is a natural coordinate on the universal covering of $S^1$, 
considered as a multivalued function on $S^1$. Note that here, as well 
as in the case of $\log \part$, the formal symbol $x$ is not an element of 
${{\tenfm G}}$ but a commutator $[x, A]$ is. 
\end{rem}

\subsection{Derivations of the algebra of $q$-pseudodifferential operators}

Analogously to the classical case, described in Section 4.1 we define here 
derivations of the deformed algebra 
${{\tenfm G}}_q$ by the symbols $\log D_q$ and $\log x$.

\subsubsection{The derivation $\protect\log D_q$.}
The classical case again gives us the recipe how to define an action of the
operator $\log D_q$ on $q$-pseudodifferential symbols
$A(x, D_q) \in {{\tenfm G}}_q$. We consider the corresponding Lie
group of $q$-symbols and its tangent structure.

\begin{prop}\label{logDqprop} 
$q$-pseudodifferential symbols of the form 
$u= D_q^\alpha + \sum_{k\ge1} u_k(x) D_q^{\alpha-k}$ constitute
an infinite dimensional Lie group $G_{q}$ acting by automorphisms on the
associative algebra $\psi DO_q$, $A\mapsto uAu^{-1}$, $u\in G_q$. The 
corresponding tangent Lie algebra is generated by integral $q$-operators 
and a symbol $\log D_q$:
${\frak G}_{q} = \{ \sum_{j=-\infty}^{-1} a_{j}(x)D_q^{j} +
\lambda \protect\log D_q \}$,
where the commutation relations for $\Psi DO_q$ were defined above and 
the commutator of $\log D_q$ with a symbol  
$ A = 
x^n D_q^p$ is
\begin{align} \label{equ3,}   
[ \log D_q, x^n D_q^p] &\overset{\text{def}}= (\log q) n x^n D_q^p 
+ \sum_{k\geq1} \frac{(-1)^{k-1}}k (\tau^{-k} (D_q^k x^n)) D_q^{p-k} \\
&= (\log q) n x^n D_q^p + \sum_{k\geq1} \frac{(-1)^{k-1}}k 
q^{-kn} (n)_q^k x^n D_q^{p-k}  \notag
\end{align}
\end{prop}

\begin{rem} Analogously to the classical case
the algebra of integral operators is closed under the commutations
with $\log D_q$. The main and crucial difference with the classics is
the presence of the knot term ($k=0$) in the formula for the $\log D_q$.
This shows that the group $G_q$ of $q$-analogs is ``solvable'' and its
properties are much different from the 
``unipotent'' $G_{Int}$ in the classical case.
\end{rem}
 
\begin{pf*}{Proof of \propref{logDqprop}}
Indeed,
\begin{align*}
\lim_{\eps\to0} \frac{D_q^\eps(u D_q^p)D_q^{-\eps} - u D^p_q} {\eps} &= 
 \lim_{\eps\to0} \frac1{\eps} \left\{ \sum_{k\ge0} \binom \eps k  
(\tau^{\eps-k} (D_q^k u)) D_q^{p-k} - u D_q^p \right\}  \\
&= \lim_{\eps\to0} \frac{\tau^\eps u-u}{\eps} D_q^p
+ \sum_{k\geq1} \frac{(-1)^{k-1}}k (\tau^{-k} (D_q^k x^n)) D_q^{p-k} .
\end{align*}
\end{pf*}

The formula for the action of $\log D_q$ can be written also as the  
following version of the Leibnitz formula:
\begin{equation} \label{logd}
[\log D_q, A] = (\log q)n A + \sum_{k\geq1} \frac{1}{k!} 
\left(\frac{d^{k-1}}{dD_q^{k-1}} D_q^{-1}\right) * \left(D_q^k A\right) 
\end{equation} 
where $A= x^nD_q$. Note that  the two logarithmic terms  do not 
cancel in \eqref{logd} as it was  in the classical case (see \ncite{K-K}), 
but generate the first right hand side term.

\begin{prop} \label{logAB}
The action of $\log D_q$ is a derivation of the associative algebra 
$\psi DO_q$:
\[ [\log D_q, A\circ B] = A\circ [\log D_q, B] + [\log D_q, A] \circ B .\]
\end{prop}

\begin{pf} This immediately follows from the fact that $\log  D_q$
is a tangent vector to the Lie group $G_q$.
\end{pf}

The inner product allows to define a 2-cocycle on the corresponding Lie 
algebra:
\[ c_D(A,B) = \int\res ([\log D_q, A] BE ) .\]

\begin{prop}
The 2-cochain $c_D(A,B)$ is antisymmetric and satisfies the 
cyclic cocycle identity.
\end{prop}

\begin{pf} First note that $\int\res ([\log D_q, A]) =0$ for 
any $A\in \G_q$ by \eqref{equ3,}. Indeed, $\int nx^n=0$ for all $n\in\Z$
and $\int\tau^{-k}(D_q^kf) =0$ for all $k>0$. This implies
\begin{align*}
0= \int\res ([\log D_q, ABE]) &= 
\int\res ([\log D_q, A] BE) + \int\res (A [\log D_q, B]E)  \\
&= c_D(A,B) + c_D(B,A)
\end{align*}
because $[\log D_q ,E] =0$ and by \thmref{res(ABE)}.
The cyclic cocycle property follows from antisymmetry identity and 
\propref{logAB}. 
\end{pf}

\begin{rem}
The restriction of this Lie 2-cocycle to the Lie algebra of 
$q$-differential operators (containing polynomials in $\part_q$ only) 
gives a non-trivial cocycle,  cohomologous to the $q$-deformation 
of the Kac-Peterson cocycle \cite{[K-P]} constructed by Kassel \cite{[Ka]}.
\end{rem}

\begin{prop}\label{Tr_D} 
In the Kac-Radul construction (see \secref{KaRa}), if
 one chooses 
\[ \Tr_D(\sum_{p=-\infty}^N a_p\xi^p)= 
\sum_{p=0}^N a_p {(-1)^p\over (q-1)^{p+1}} 
\Big\{ \log q + \sum_{k=1}^p  {(1-q)^k\over k} \Big\} \]
as the trace on $A$, then the 2-cocycle $\psi_{\s , \Tr}$ induces
on $\G _q$ the same cocycle $c_D$  described above.
\end{prop}

\begin{pf} Compute explicitly the cocycle $c_D$:
\begin{align*}
c_D(x^m\s^m(D_q^a), x^nD_q^b) &= 
\int\res(x^nD_q^b \circ [\log D_q,x^m] \s^m(D_q)^a \circ E) \\
= \d_{n+m,0} & \res \Bigl\{\s^m(D_q)^{a+b} \bigl(m\log q + 
\sum_{k>0} \frac{(-1)^{k-1}}k q^{-km} (m)_q^k D_q^{-k} \bigr) E\Bigr\} .
\end{align*}
On the other hand
\[ \psi(x^m\s^m(\xi^a), x^n\xi^b) = 
\d_{n+m,0} \Tr_D(1+\s+\dots+\s^{m-1}) \xi^{a+b} \]
for $m>0$. This gives an equation to determine $\Tr_D$
\be\label{TrsigD}
\nquad \Tr_D(1+\s+\dots+\s^{m-1}) \xi^c = \res \bigl\{\s^m(\xi)^c 
\bigl(m\log q + \log(1+q^{-m}(m)_q\xi^{-1}) \bigr) E\bigr\} .
\end{equation}
Put $m=1$ and define $\Tr_D$ via the right hand side
\[ \Tr_D\xi^c = \res \Bigl\{ (q\xi+1)^c 
\bigl(\log q+ \sum_{k>0} \frac{(-1)^{k-1}}k q^{-k}\xi^{-k} \bigr)
((q-1)\xi+1)^{-1} \Bigr\} \]
in particular, $\Tr_D\xi^c=0$ if $c<0$.

We want to prove \eqref{TrsigD} for $m>1$. Summing up these equations
with weights $\sum_{c=0}^b \binom bc (q-1)^c$ we get an equivalent form
\[ \Tr_D(1+\s+\dots+\s^{m-1})(1+(q-1)D)^b =
\res \bigl\{ (1+(q-1)\s^m(\xi))^b \log(q^m+(m)_q\xi^{-1}) E\bigr\} .\]
Since $\s(1+(q-1)\xi) = q(1+(q-1)\xi)$ this is the same as
\[ (m)_{q^b} \Tr_D(1+(q-1)\xi)^b = q^{mb}
\res \bigl\{ (1+(q-1)\xi)^{b-1} \log(q^m+(m)_q\xi^{-1}) \bigr\} .\]
The definition of $\Tr_D$ gives
\begin{align*}
\Tr_D(1+(q-1)\xi)^b &= 
\res \bigl\{ (1+(q-1)(q\xi+1))^b \log(q+\xi^{-1}) E\bigr\} \\
&= q^b \res \bigl\{ (1+(q-1)\xi)^{b-1} \log(q+\xi^{-1}) \bigr\} .
\end{align*}
So the identity to prove is
\[ (m)_{q^{-b}} \res \bigl\{ (1+(q-1)\xi)^{b-1} \log(q+\xi^{-1}) \bigr\} =
\res \bigl\{ (1+(q-1)\xi)^{b-1} \log(q^m+(m)_q\xi^{-1}) \bigr\} \]
for $b\in\Z_{>0}$. Computing the residue by differentiation with respect
to $q^{-m}$ we see that the both sides are equal to 
$b^{-1}(q^{-mb}-1)(1-q)^{-1}$. This gives a simple expression for $\Tr_D$
\begin{align*}
\Tr_D(1+(q-1)\xi)^b &= \frac{(b)_q}b \qquad \text{for }b>0, \\
\Tr_D1 &= \frac{\log\ q}{q-1} ,
\end{align*}
whence
\begin{align} 
\Tr_D\xi^p &= (-1)^p{\log q\over (q-1)^{p+1}} + {1\over(q-1)^p} 
\sum_{k=1}^p (-1)^{p-k} \binom pk {(k)_q\over k} \notag \\
&= {(-1)^p \over (q-1)^{p+1}} \Bigl( \log q + \sum_{k=1}^p (-1)^k 
\binom pk \bigl(\frac{q^k}k - \frac1k \bigr) \Bigr) \label{TrDxi}
\end{align}
for $p\in\Z_{\ge0}$. Denote 
$f(q) = \sum_{k=1}^p (-1)^k \binom pk \frac{q^k}k$. Since 
$qf'(q) = (1-q)^p-1$ we can write $f'(q) = - \sum_{n=0}^{p-1} (1-q)^n$.
Plugging $f(q)-f(1) = \sum_{n=1}^p \frac{(1-q)^n}n$ to \eqref{TrDxi} we
obtain
\[ \Tr_D\xi^p = {(-1)^p \over (q-1)^{p+1}} 
\Bigl( \log q + \sum_{k=1}^p  {(1-q)^k\over k} \Bigr) .\]  
\end{pf}

\begin{rem}\label{qto1TrD}
For $q$ close to 1 and $p\ge0$ this trace can be presented as
\[ \Tr_D \xi^p = \sum_{m=0}^\infty \frac{(1-q)^m}{m+p+1} .\]
In particular, when $q\to1$ the trace $\Tr_D$ is non-singular and 
$\lim_{q\to1} \Tr_D\xi^p = {1\over p+1}$.
\end{rem}

\subsubsection{The derivation $\protect\log x$.}
Analogously, one can define a twisted derivation $l_x$:
$$l_x(f D_q^a) = a f(x) D_q^{a - 1}= {d \over dD_q}  (f D_q^a)$$
for any $A \in {\psi DO}_q$.

\begin{prop}
 The map $l_x$ is a twisted derivation of the associative algebra
${\psi DO}_q$:
\[ l_x(AB) = l_x(A) \tau(B) + A\cdot l_x(B) \]
\end{prop}

\begin{pf} Calculation. \end{pf}

\begin{cor} The map $[\log x,-]$ determined by
\[ [\log x,A] = - l_x(A) E^{-1} = - l_x(A) ((q-1)D_q+1) \]
is a derivation of the associative algebra  ${\psi DO}_q$.
\end{cor}

\begin{pf} Indeed,
\begin{align*}
[\log x,AB] &= - l_x(AB) E^{-1} \\
&= - l_x(A) E^{-1}BE E^{-1} - A\cdot l_x(B) E^{-1} \\
&= [\log x,A] B + A [\log x, B]  .
\end{align*}
\end{pf}

\begin{prop}
 The bilinear form 
\[c_x (A, B) = \int \res (A [\log x,B]E) = - \int \res (A\, l_x(B))\]
 is a skew-symmetric cyclic cocycle.
\end{prop}

\begin{pf}
Indeed,
\begin{align*}
0= \int\res (l_x( AB)) &= 
\int\res (l_x(A) \tau(B)) + \int\res ( A\, l_x(B) )  \\
&= \int\res ( B\, l_x(A) ) + \int\res ( A\, l_x(B) )
\end{align*}
implies that $c_x$ is skew-symmetric. Also,
\begin{align*}
c_x(A,BC) &= - \int\res (AB\, l_x(C)) - \int\res ( A\, l_x(B) \tau(C))  \\
&= c_x(AB,C) + c_x(CA,B)
\end{align*}
is the cyclic cocycle property.
\end{pf}

When restricted to differential operators, the cocycle $c_x$ vanishes.

\begin{rem}
In the case of the algebra $C^\infty(S^1)$ we can introduce a coordinate 
$\t$ on the universal covering $\RR^1$ of the circle $S^1$ so that 
$x = e^{i \t}$. This is not a function from $C^\infty(S^1)$, but for 
any $A \in {{\tenfm G}}_q$ we can form a commutator 
$[ \t,A] \in {{\tenfm G}}_q$. 
\end{rem}

Now we find the commutator of the two derivations.

\begin{prop}\label{[log,log]}
For any $A\in\G_q$ 
\[ [\log D_q, [\log x, A]] - [\log x, [\log D_q, A]] = [D_q^{-1} ,A] .\]
\end{prop}

\begin{pf}
Assume $A=x^n D_q^a$. The left hand side is
\begin{align*}
 [\log D_q, & [\log x, x^n D_q^a]] - [\log x, [\log D_q, x^n D_q^a]] = \\
&= -a \Bigl\{ (\log q) n x^n + \sum_{k\geq1} \frac{(-1)^{k-1}}k 
q^{-kn} (n)_q^k x^n D_q^{-k}  \Bigr\} (D_q^{a-1} + (q-1)D_q^a) \\
&\ + a(\log q) n x^n (D_q^{a-1}+ (q-1)D_q^a) \\
&\ + \sum_{k\geq1} \frac{(-1)^{k-1}}k q^{-kn} (n)_q^k (a-k) x^n 
(D_q^{a-k-1} + (q-1)D_q^{a-k}) \\
&= x^n (D_q^{a-1} + (q-1)D_q^a) 
\sum_{k\geq1} (-1)^k q^{-kn} (n)_q^k D_q^{-k} \\
&= - (n)_q x^n (D_q^{a-1} + (q-1)D_q^a) (q^nD_q + (n)_q)^{-1} . 
\end{align*}

The right hand side is
\begin{align*}
[ D_q^{-1}, x^n D_q^a] &= x^n (q^nD_q + (n)_q)^{-1} D_q^a - x^n D_q^{a-1} \\
  &= x^n (q^nD_q + (n)_q)^{-1} D_q^{a-1}((1 - q^n) D_q - (n)_q) \\
  &= (n)_q x^n (q^nD_q + (n)_q)^{-1} D_q^{a-1}((1 - q) D_q - 1) 
\end{align*}
which coincides with the expression above.
\end{pf}

\begin{cor}
One can equip the vector space 
$\bar{\G}_q = \G_q\oplus\C\log D_q \oplus \C\log x$ with a Lie algebra 
structure such that $\G_q$ is an ideal and
\[ [\log D_q, \log x] = D_q^{-1} .\]
There exists an exact sequence of Lie algebras 
$0\to \G_q\to \bar{\G}_q \to \C^2 \to0$.
\end{cor}

\subsection{$3$-cocycles on $\G $ and $\G _q$.}
Since the Lie algebra $\G_q$ has an invariant form it inherits also the 
standard 3-cocycle $c_0^q(L,M,N)  = \int\res([L,M]NE)$ defined for $q\ne1,0$.
For $q=1$ there is a cocycle $c_0^1(L,M,N)  = \int\res([L,M]N)$ on $\G$.

\begin{prop}\label{proqto1}
The cocycle $c_0^q$ is non-singular as $q\to1$ and, moreover,
\[ \lim_{q\to1} c_0^q = c_0^1 .\]
\end{prop}

\begin{pf}
It suffices to check this for $L=x^lD_q^a$, $M=x^mD_q^b$, 
$N=D_q^c\circ x^n$. Change for $q\ne1$ the variable $D_q$ to $T=(q-1)D_q+1$,
then $\psi DO_q$ consists of formal power series 
$\sum_{k=-\infty}^N u_k(x)T^k$. Define $\res_T: \psi DO_q\to F$,
$\res_T\bigl( \sum_{k=-\infty}^N u_k(x)T^k \bigr) = u_{-1}(x)$, then
$\res_T = (q-1)\res$. Substituting $D_q = T(1-T^{-1})/(q-1)$ we find
{\allowdisplaybreaks
\begin{align*}
\int&\res([L,M]NE) = \frac1{q-1} \int\res_T([L,M]NT^{-1}) \\
&= \frac1{(q-1)^{a+b+c+1}} \int\res_T\bigl\{ x^n[x^lT^a(1-T^{-1})^a, 
x^mT^b(1-T^{-1})^b] T^c(1-T^{-1})^cT^{-1} \bigr\} \\
&= \frac1{(q-1)^{a+b+c+1}} \int\res_T\bigl\{ 
x^{l+m+n} \bigl( (q^mT)^a (1-q^{-m}T^{-1})^a T^b (1-T^{-1})^b \\*
&\hspace{4cm} - (q^lT)^b (1-q^{-l}T^{-1})^b T^a (1-T^{-1})^a
\bigr) T^{c-1}(1-T^{-1})^c \bigr\} \\
&= \frac{\delta_{l+m+n,0}}{(q-1)^{a+b+c+1}} \res_T\bigl\{ q^{ma}T^{a+b+c-1}
\sum_{k\ge0}(-1)^k\binom ak q^{-mk}T^{-k} \cdot 
\sum_{p\ge0}(-1)^p\binom{b+c}p T^{-p} \\*
&\hspace{4cm} - q^{lb}T^{a+b+c+1} \sum_{k\ge0}(-1)^k\binom bk q^{-lk}T^{-k}
\cdot \sum_{p\ge0}(-1)^p\binom{a+c}p T^{-p} \bigr\} \\
&= (-1)^{a+b+c} \frac{\delta_{l+m+n,0}}{(q-1)^{a+b+c+1}} \bigl\{
\sum\begin{Sb}k+p=a+b+c \\ k,p\ge0\end{Sb} 
\binom ak \binom{b+c}p q^{m(a-k)} \\*
&\hspace{6cm} -\sum\begin{Sb}k+p=a+b+c \\ k,p\ge0\end{Sb} 
\binom bk\binom{a+c}p q^{l(b-k)} \bigr\}
\end{align*}
}

For $s\in\Z_{\ge0}$, $a\in\Z$ introduce the function
\[ f_a(z) = \sum\begin{Sb}k+p=s \\ k,p\ge0\end{Sb} 
\binom ak\binom{s-a}p z^{a-k} .\]
It is the coefficient at $t^s$ in the decomposition
\be\label{(1+t)} 
(z+t)^a(1+t)^{s-a} = \sum_{m\ge0} t^m 
\sum\begin{Sb}k+p=s \\ k,p\ge0\end{Sb} \binom ak\binom{s-a}p z^{a-k} .
\end{equation}
Differentiating the last relation with respect to $z$ we find
\begin{align*}
f_a(1) &= 1, \\
\frac{d^n}{dz^n} f_a(z) \big|_{z=1} &= n! \binom an \binom{s-n}s ,
\end{align*}
in particular, $f_a^{(n)}(1) =0$ for $1\le n\le s$. Therefore,
$f_a(z) = 1 + (z-1)^{s+1}g_a(z)$ for some Laurent polynomial $g_a(z)$,
such that $g_a(1) = (-1)^s \binom a{s+1}$.

Assuming $s=a+b+c\ge0$ and plugging $f_a(q^m) - f_b(q^l)$ to $c_0^q(L,M,N)$
we see that the limit exists and
\[ \lim_{q\to1} \int\res([L,M]NE) = \delta_{l+m+n,0} \bigl\{
\binom a{a+b+c+1} m^{a+b+c+1} - \binom b{a+b+c+1} l^{a+b+c+1} \bigr\} .\]

Now we calculate the value of $c_0^1$ on $L=x^lD^a$, $M=x^mD^b$, 
$N=D^c \circ x^n$ when $q=1$:
\begin{align*}
\int\res([L,M]N) &= \int\res( [x^lD^a, x^mD^b] D^cx^n ) \\
&= \int\res\bigl\{ x^{l+m+n}((D+m)^aD^b - (D+l)^bD^a) D^c \bigr\} \\
&= \delta_{l+m+n,0} \res\bigl\{ \sum_{k\ge0} \binom ak m^k D^{a-k+b+c}
- \sum_{k\ge0} \binom bk l^k D^{b-k+a+c} \bigr\} \\
&= \delta_{l+m+n,0} \bigl\{
\binom a{a+b+c+1} m^{a+b+c+1} - \binom b{a+b+c+1} l^{a+b+c+1} \bigr\}
\end{align*}
which coincides with the previous limit.
\end{pf}

\begin{rem}\label{remc|DO}
If $a,b,a+c,b+c\ge0$, then $c_0^q(x^lD_q^a, x^mD_q^b, D_q^c\circ x^n) =0$.
Indeed, as follows from \eqref{(1+t)} $f_a(z)=1$ for $0\le a\le s$. 
In particular, the restriction of $c_0^q$ to the subalgebra $DO_q$
 vanishes identically.
\end{rem}

Two outer derivations $\xi=\ad{\log D_q}$ and 
$\eta=\ad{\log x} \in H^1 (\G _q, \G _q) $ produce not only two 
complex valued two-cocycles. They also induce a 2-cocycle 
$f = \eta \wedge \xi= m(\eta\boxtimes\xi - \xi\boxtimes\eta) $ in $ Z^2 
(\G _q, \G _q)$: it can be considered as the cup product
of those two classes. 

\begin{rem}
In contrast to the classical case of Poisson algebra, the cocycle $f$ is
non-trivial, at least for $q=1$, as the following argument shows.
Since $\G$ is isomorphic to its own dual as a representation of itself, one 
has $f\in H^2(\G,\G^*)$ and one can check that its restriction to vector 
fields $\A(S^1)$ is non-trivial: one has $I^*(f) \in H^2(\A(S^1),\G^*)$ but 
as an $\A(S^1)$-module $\G^*$ splits into a direct sum 
$\G^*=\oplus_{n\in\Z}\F_n$, where $\F_n$ is the space of $n$-densities. Then
$I^*(f) \in H^2(\A(S^1,\F_2)$ and one computes easily 
\[ I^*(f)(a\frac{\part}{\part x} , b\frac{\part}{\part x}) 
= (a''b'-a'b'') dx^2 .\]
The non-triviality of the cohomology class of this cocycle can either be
proved directly using Fuchs' computations \cite{[Fu]} for non trivial
representations of $\A(S^1)$ or either remark that $\F_2=\A(S^1)^*$ and
this class $I^*(f)$ then induces the Godbillon-Vey generator in 
$H^3(\A(S^1),\RR)$.
\end{rem}

The cocycle $f$ will give rise to a non trivial complex valued 
three-cocycle on $\G _q$ via the  map $\theta$.

\begin{thm}\label{CGV3} 
The complex valued cochain on $\G _q$  defined by 
\begin{multline*} 
c_{GV}(L, M, N) = \Alt_{ L, M, N }
\int \res \bigl(( \xi(M)N\eta(L) - M\xi(N)\eta(L) - LMND_q^{-1})E \bigr) \\
= \Alt_{L,M,N} \int \res  \bigl(M\xi( N)l_x(L) 
- \xi(M)Nl_x(L) - LMND_q^{-1} + (q-1)LMNE \bigr) 
\end{multline*}
is a three-cocycle, defined for $q\ne0,1$ and having a limit as $q\to1$ 
which is a non trivial cocycle on $\G $
\[ c_{GV}(L, M, N) = \Alt_{ L, M, N }
\int \res \bigl( \xi(M)N\eta(L) - M\xi(N)\eta(L) - LMND^{-1} \bigr) .\]
\end{thm}

\begin{pf} Consider the 3-cochain
\[ \t(f)(L,M,N) = \Alt_{L,M,N} \Tr(\eta(L)\xi(M)N - \xi(L)\eta(M)N) \]
(sum over 6 permutations). Its coboundary is given by the \lemref{dthetalem}
\begin{alignat*}2
d\t(f)(L,M,N,P) &= \Alt_{L,M,N,P}\Tr(\eta(L)\xi(M)[N,P] -\xi(L)\eta(M)[N,P])
&&\quad \text{(12 shuffles)} \\
&= \Alt_{L,M,N,P}\Tr(\eta(L)\xi(M)NP -\xi(L)\eta(M)NP)
&&\nquad \text{(24 permutations)} \\
&\overset{(1)}=  \frac12 \Alt_{L,M,N,P}\Tr([\eta,\xi](L)MNP)
&&\nquad \text{(24 permutations)} \\
&= - \frac12 \Alt_{L,M,N,P}\Tr([D_q^{-1},L]MNP) 
&&\nquad \text{(24 permutations)} \\
&= - \Alt_{L,M,N,P}\Tr(D_q^{-1}LMNP) 
&&\nquad \text{(24 permutations)} \\
&\overset{(2)}= - da(L,M,N,P) 
\end{alignat*}
where 
\[ a(L,M,N) = \Alt_{L,M,N}\Tr(D_q^{-1}LMN) 
\qquad\text{ (sum over 6 permutations).} \]
Here we used \propref{[log,log]} and equation (1) is deduced by
``integration by parts'', that is from the property $\Tr(\xi(A))=0$, 
$\Tr(\eta(A))=0$ for any $A\in\G_q$. Equation (2) illustrate the 
difference between odd and even cochains.

This implies that
\[ c_{GV}(L,M,N) = \t(f)(L,M,N) + a(L,M,N) \]
is a 3-cocycle well defined for $q\ne0,1$. The formula
\[ c_{GV}(L,M,N) = \Alt_{L,M,N} \int \res  \bigl(M\xi( N)l_x(L) 
- \xi(M)Nl_x(L) - LMND_q^{-1} ((q-1)D_q+1)^{-1} \bigr) \]
shows that the cocycle has a limit as $q\to1$ iff its last summand
$$\Alt_{L,M,N} \int \res  ( LMND_q^{-1} ((q-1)D_q+1)^{-1} )$$ 
has a limit. Since
\be\label{frafra} 
\frac1{D_q((q-1)D_q+1)} = \frac1{D_q} - \frac{q-1}{(q-1)D_q+1} 
\end{equation}
we have
\begin{align*}
\Alt_{L,M,N} \int \res ( LMND_q^{-1}E) &= 
\Alt_{L,M,N} \int\res (LMND_q^{-1}) - (q-1) \Alt_{L,M,N} \int\res (LMNE) \\
&= \Alt_{L,M,N} \int\res (LMND_q^{-1}) - 3(q-1) c_0^q(L,M,N) .
\end{align*}
We know that $c_0^q$ is not singular when $q$ tends to 1 by 
\propref{proqto1}. Therefore, 
\begin{multline*} 
c_{GV}(L,M,N) = \\ 
= \Alt_{L,M,N} \int \res  \bigl(M\xi( N)l_x(L) 
- \xi(M)Nl_x(L) - LMND_q^{-1} \bigr) + 3(q-1) c_0^q(L,M,N) 
\end{multline*}
is not singular either and
\[ \lim_{q\to1} c_{GV}(L,M,N) = \Alt_{L,M,N} \int \res \bigl(M\xi( N)l_x(L)
- \xi(M)Nl_x(L) - LMND^{-1}  \bigr) .\]
The last cocycle on $\G$ can be obtained also via the same reasoning
as for $q\ne1$.

Non-triviality of the cocycle for $q=1$ follows from explicit calculation. 
Computing it on the first order difference operators 
$L=x^lD_q$, $M=x^mD_q$, $N=x^nD_q$ we get
\[ c_{GV}(x^lD_q,x^mD_q,x^nD_q) = 
\frac12 \d_{l+m+n,0} \Alt_{l,m,n} q^{-n}(m)_q(n)_q^2 .\]
As $q\to1$ the operator $D_q$ tends to $D=x\part_x$ and $L,M,N$ tend to 
vector fields on $S^1$. Therefore, the restriction of this cocycle at $q=1$ 
to the algebra of vector fields $\A(S^1)$ is
\[ c_{GV}(fD,gD,hD) = \frac12 \int_{S^1}
\left\vert{ 
\begin{matrix}
f&Df&D^2f\\
g&Dg&D^2g \\
h&Dh&D^2h
\end{matrix}
 }\right\vert
\frac{dx}x \] 
which is the generator of the $H^3 (\A(S^1), \CC)$ found in \cite{[G-Fu]}.
\end{pf}

The Godbillon-Vey form of this cocycle after restriction to vector fields
suggests the name ``quantum'' Godbillon-Vey $3$-cocycle for $c_{GV}(L,M,N)$.

\begin{prop} 
The cocycle $c_{GV}$ extends to a 3-cocycle on the Lie
algebra $\G _q \oplus \R\log D_q$ by the formula
\[ c_{GV}(\log D_q,L,M) = 
3\int\res \bigl\{( L[\log D_q,M]D_q^{-1} + LD_q^{-1}[\log D_q,M] )E\bigr\}\]
for $L,M\in \psi DO_q$, $q\ne0,1$. When $q\to1$ the cocycle is not
singular and
\[ \lim_{q\to1} c_{GV}(\log D_q,L,M) = 
3\int\res \bigl\{ L[\log D,M]D^{-1} + LD^{-1}[\log D,M] \bigr\} .\]
\end{prop}

\begin{pf}
Since $\Tr(D_q^\alpha L D_q^{-\alpha}) = \Tr L$ for $\alpha\in\R$,
\begin{align*}
c_{GV}([\log D_q,L],M,N) &+ c_{GV}(L,[\log D_q,M],N) + 
c_{GV}(L,M,[\log D_q,N]) = \\
&= \frac d{d\alpha} c_{GV}(D_q^\alpha L D_q^{-\alpha}, 
D_q^\alpha M D_q^{-\alpha}, D_q^\alpha N D_q^{-\alpha}) \big|_{\alpha=0} \\
&= \Alt_{L,M,N} \Tr ((\xi(M)N - M\xi(N)) [\eta,\xi](L)) \\
&= \Alt_{L,M,N} \Tr(\xi(L)[D_q^{-1},M]N - [D_q^{-1},L]\xi(M)N) \\
&= da_2(L,M,N)
\end{align*}
the cochain $a_2$ being defined as
\[ a_2(L,M) = 3\Tr(\xi(L)D_q^{-1} M + D_q^{-1} \xi(L)M) \]
(we used the obvious $\xi(D_q^{-1})=0$). Therefore, extending $c_{GV}$
to a cochain on $\G _q \oplus \R\log D_q$ by
$c_{GV}(\log D_q,L,M) = - a_2(L,M)$ we obtain a cocycle.

Using again \eqref{frafra} we transform the cocycle to
\begin{align*} 
c_{GV}(\log D_q,L,M) &= 3\int\res \bigl\{( L[\log D_q,M] + [\log D_q,M]L)
(D_q^{-1} - (q-1)E) \bigr\} \\
&= 3\int\res \bigl\{( L[\log D_q,M] + [\log D_q,M]L)D_q^{-1} \bigr\}
- 6(q-1) c_D(M,L) .
\end{align*}
This is non-singular because $c_D$ is not singular by \propref{Tr_D},
whence the formula for limit follows.
\end{pf}

\subsection{Higher cocycles on $\G$ and $\G_q$.}
Gelfand and Mathieu constructed a family of Lie algebra cocycles in the
following situation. Let $A$ be an associative algebra and $\frak h$ be
an abelian Lie subalgebra of $\Der A$. Let $\Tr :A\to \C$  be 
$\frak h$-invariant trace. Then there are some cocycles from 
$Z^k((A,[,]),\C)$ associated with these data (see \cite{[G-M]}). We 
consider similar case of 2-dimensional subspace $\frak h\subset \Der A$
such that its image in $\operatorname{Out}A = \Der A / \operatorname{Inn}A$
is an abelian Lie subalgebra. That is, we consider two derivations 
$\xi,\eta$ such that their commutator is inner. The results are formulated
in the particular case $A=\psi DO_q$, $\xi=\ad\log D_q$, $\eta=\ad\log x$.

\begin{prop} 
For any odd $n\ge1$ there is a cocycle in $Z^n(\G_q,\C)$ 
\[ c^{(n)}(X_1,\dots,X_n) = \Alt_{X_1,\dots,X_n} \Tr(X_1\cdots X_n).\]
For $n\ge3$ it is continuous at $q=1$:
\[ \lim_{q\to1} c^{(n)} (X_1,\dots,X_n) = c^{(n)}\big|_{q=1}
(\lim_{q\to1} X_1,\dots,\lim_{q\to1} X_n) .\]                
For $n\ge3$ (and for $n=1$ if $q=1$) its restriction to $DO_q$ vanishes.
\end{prop}

\begin{pf} The first claim is standard (cf. \cite{[G-M]}). The other claims
follow from a particular case $n=3$, considered in \propref{proqto1}
and Remark~\ref{remc|DO}. 
\end{pf}

\begin{prop} 
For any even $n\ge2$ there are cocycles in $Z^n(\G_q,\C)$ 
\begin{align*} 
c^{(n)}_D (X_1,\dots,X_n) &= 
\Alt_{X_1,\dots,X_n} \Tr([\log D_q, X_1] X_2\cdots X_n),\\
c^{(n)}_x (X_1,\dots,X_n) &= 
\Alt_{X_1,\dots,X_n} \Tr([\log x, X_1] X_2\cdots X_n).
\end{align*}
They are continuous at $q=1$:
\[ \lim_{q\to1} c^{(n)}_D  = c^{(n)}_D\big|_{q=1}, \qquad
\lim_{q\to1} c^{(n)}_x  = c^{(n)}_x\big|_{q=1} .\]
The restriction of $c_x^{(n)}$ to $DO_q$ vanishes.
\end{prop}

\begin{pf} The first claim follows from results of Gelfand and Mathieu 
\cite{[G-M]}. Continuity of $c_D$ follows from the particular case $n=2$
(see \propref{Tr_D} and Remark~\ref{qto1TrD}). The claims about $c_x$
follow from the presentation 
\[ c^{(n)}_x (X_1,\dots,X_n) = 
- \Alt_{X_1,\dots,X_n} \int\res( X_2\cdots X_n l_x(X_1)) \]
valid for any $q$. 
\end{pf}

\begin{thm} 
For any odd $n\ge3$ there is a cocycle in $Z^n(\G_q,\C)$ 
\begin{multline*} 
c^{(n)}_{x,D} (X_1,\dots,X_n) = \Alt_{X_1,\dots,X_n} 
\Tr\bigl(\sum_{i=1}^{\frac{n-1}2} \xi(X_1) X_2\cdots\eta(X_{2i})\cdots X_n\\
- \xi(X_1) X_2\cdots  X_{n-1} \eta(X_n) - D_q^{-1}X_1 X_2\cdots X_n\bigr).
\end{multline*}
It is continuous at $q=1$:
\[ \lim_{q\to1} c^{(n)}_{x,D} (X_1,\dots,X_n) = c^{(n)}_{x,D}\big|_{q=1}
(\lim_{q\to1} X_1,\dots,\lim_{q\to1} X_n) .\]                
\end{thm}

\begin{pf} Consider the $n$-cochain
\[ b_n (X_1,\dots,X_n) = \Alt_{X_1,\dots,X_n} \Tr\bigl(
\sum_{i=1}^{\frac{n-1}2} \xi(X_1) X_2\cdots\eta(X_{2i})\cdots X_n
- \xi(X_1) X_2\cdots  X_{n-1} \eta(X_n) \bigr) \]
which can be written as
\[ b_n = \Tr\bigl( \sum_{i=1}^{\frac{n-1}2} 
\xi \cup m^{2(i-1)} \cup\eta\cup m^{n-2i} 
- \xi \cup m^{n-2} \cup \eta \bigr) ,\]
where $m^k(Y_1,\dots,Y_k) = \Alt_{Y_1,\dots,Y_k} Y_1\cdots Y_k$
and $a\cup b \overset{\text{def}}= \mu(a\boxtimes b)$, $\mu$ being the
multiplication.
Since $m^k=m^1\cup m^1\cup\dots\cup m^1$ and $dm^1=m^2$, we have
$dm^k=m^{k+1}$ for odd $k$, and $dm^k=0$ for even $k$. Therefore,
\begin{align*}
db_n &= \Tr\bigl(\sum_{i=1}^{\frac{n-1}2} 
\xi\cup m^{2(i-1)}\cup\eta\cup m^{n+1-2i} + \xi\cup m^{n-1}\cup\eta\bigr) \\
&= \Tr\sum_{i=1}^{\frac{n+1}2} \xi\cup m^{2(i-1)}\cup\eta\cup m^{n+1-2i} \\
&= \frac12 \Tr([\xi,\eta]\cup m^n)
\end{align*}
due to the following identity
\begin{align*} 
\frac12 & \Alt\Tr\bigl( (\xi\eta)(X_1)X_2\cdots X_{n+1}
- (\eta\xi)(X_1)X_2\cdots X_{n+1} \bigr) = \\
&= \frac12 \Alt\Tr\bigl( 
-\sum_{i=2}^{n+1} \eta(X_1) X_2\cdots\xi(X_i)\cdots X_{n+1}
+\sum_{j=2}^{n+1} \xi(X_1) X_2\cdots\eta(X_j)\cdots X_{n+1} \bigr) \\
&= \frac12 \Alt\Tr\bigl( 
-\sum_{i=2}^{n+1} (-1)^{i-1}\xi(X_1)X_2\cdots\eta(X_{n-i+3})\cdots X_{n+1}\\
&\hspace{8cm} 
+\sum_{j=2}^{n+1} \xi(X_1) X_2\cdots\eta(X_j)\cdots X_{n+1} \bigr) \\
&= \frac12 \Alt\Tr \sum_{j=2}^{n+1} [(-1)^{n+3-j}+1] 
\xi(X_1) X_2\cdots\eta(X_j)\cdots X_{n+1} \\
&= \Alt\Tr \sum_{i=1}^{\frac{n+1}2} 
\xi(X_1) X_2\cdots\eta(X_{2i})\cdots X_{n+1} .
\end{align*}
Hence,
\[ db_n = \frac12 \Tr(\ad D_q^{-1} \cup m^n) \]
and we get
\begin{align*}
db_n(X_1,\dots,X_{n+1}) &= \frac12 \Alt_{X_1,\dots,X_{n+1}} 
\Tr(D_q^{-1}X_1X_2\cdots X_{n+1} - X_1 D_q^{-1}X_2\cdots X_{n+1}) \\
&= \Alt_{X_1,\dots,X_{n+1}} \Tr(D_q^{-1}X_1X_2\cdots X_{n+1}) \\
&= da_n(X_1,\dots,X_{n+1}) ,
\end{align*}
where
\[ a_n(X_1,\dots,X_n) = \Alt_{X_1,\dots,X_n}\Tr(D_q^{-1}X_1X_2\cdots X_n).\]
This implies that $c_{x,D}^{(n)} = b_n - a_n$ is a cocycle.

Using the presentation
\begin{multline*} 
b_n(X_1,\dots,X_n) = \\ 
= \Alt_{X_1,\dots,X_n} \int\res\bigl(
-\sum_{i=1}^{\frac{n-1}2} X_{2i+1} \cdots X_n\xi(X_1)X_2 \cdots l_x(X_{2i})
+ \xi(X_1) X_2\cdots  X_{n-1} l_x(X_n) \bigr) 
\end{multline*}
we see that $b_n$ is continuous when $q$ tends to 1. The continuity of
$a_n$ follows from identity~\eqref{frafra} and the continuity of $c^{(n)}$
as in the proof of \thmref{CGV3}.
\end{pf}

Results of Feigin~\cite{[Fe]} give some support to the conjecture of
non-triviality of the cocycles above.

\section{Deformations of the cocycles.} \label{sec5}
One can ask how the cohomology is modified under formal deformations. So 
let ${\cal G}_t$ be a formal deformation of a Lie algebra $ {\cal G =
\cal G}_0$ and $M_t$ a formal deformation of a $ {\cal G}$ module 
$M = M_0$. In the finite dimensional case, one has the ``Fuchs principle'' 
which states that the cohomology cannot increase in dimension through 
deformation: $\dim H^* ( {\cal G}_t;M_t) \le \dim H^* ( {\cal G}_0;M_0)$. 
The Hochschild cohomology version of this result is carefully described 
in the work of M. 
Gerstenhaber and S.~
D. Schack \cite{[G-S]}.
 The authors acknowledge the priority 
 of Nijenhuis and Richardson (unpublished) for the Lie algebra cohomology
case.

In many cases, the cohomology decreases effectively, and in particular, the
cohomology class which leads the deformation gets killed in the cohomology 
of the deformation; intuitively speaking, formal or actual deformations make
cohomology simpler and simpler. But this is not always the case. We shall 
prove that the  Vey class survives in the cohomology of the Lie algebra
of pseudo-differential operators.

Let $( {\cal G}_t,M_t)$ be a formal deformation of a Lie algebra and
module $ ( {\cal G}_0,M_0)$ as above. Let
$(C^*_t,d_t)$ and $(C_0^*,d_0)$ be the corresponding Chevalley-Eilenberg
complexes; one has a one parameter family of differentials on the same
underlying graded space. Let $c_t$ be a family of cocycles: $d_t(c_t)= 0$. 
If the limit $c_0 = \lim_{t \to 0} c_t$ is a $d_0$-cocycle (i.e. 
$d_0 (c_0) = 0$)
one says that the cohomology class $[c_t]$ contracts onto $[c_0]$. 

\begin{prop}
 There exists a nontrivial
cohomology class $[\wtilde V]
\in H^2 (\G , \G )$ which contracts onto the Vey class $[V] \in H^2
(N,N)$.
\end{prop}

We shall describe the phenomenon of contraction of cohomology classes 
by using graded Lie algebra techniques as developed by P.B.A.~Lecomte~
(\cite{[Le]}). Let $E$ be the underlying vector space of the Lie algebras we 
consider, and $A^*(E)$ be the Richardson-Nijenhuis graded Lie algebra 
on $E$, whose bracket will be denoted $[[\ ,\ ]]$. The Lie algebra 
structure $ {\cal G}_0$ is then defined by some 
$c_0 \in A^1(E)$ such that $[[c_0,c_0]]=0$.

 Let 
$c_t = c_0 + \sum_{i \ge 1}^{+ \infty} t^{i} c_i$ be the formal deformation
giving the structure on $ {\cal G}_t$; and then $[[c_t,c_t]] = 0$. This
deformation being supposed to be nontrivial, the cocycle $c_1$ induces a 
nontrivial cohomology class in $H^2( {\cal G}_0, {\cal G}_0)$. One can then
construct a cocycle on $ {\cal G}_t$ whose cohomology class contracts onto
the class of $c_1$; simply take the formal series given by derivative of 
$c_t : \ \dot c_t = c_1 + \sum_{i \ge 2}^{+ \infty} i \, t^{i-1} c_i$. One 
obviously has $[[ \dot c_t,c_t]] = 0$ by derivation of $[[c_t,c_t ]] = 0$ 
and $\dot c_t$ contracts onto $c_1$. 

Let us check whether the cocycle 
$\dot c_t$ is cohomologically trivial or not for $t > 0$. If so there 
exists a formal series $a_t \in A^0 (E) [[t]]$ such that 
$\dot c_t = [[c_t,a_t]]$. But $a_t$ must be singular at $0$ since $c_0$ 
is not cohomologically trivial. So let us suppose
$a_t = {1 \over t} \left ( a_0 + \sum_{t \ge 1}^{} t^{i} a_i \right )$. 
One deduces from $\dot c_t = [[c_t,a_t ]]$ that

{(i)} $[[c_0,a_0]] = 0 $ 

{(ii)} $[[c_0,a_1]] + [[c_1,a_0]] = c_1 $
 
 \noi so (i) $a_0$ is a cocycle
and (ii) the action of $a_0$ on cohomology leaves the class of $c_1$ 
invariant (the term $[[c_0,a_1]]$ being nothing but the coboundary of $a_1$).

Straightforward  example: let $E = k^3$  ($k=\RR \text{ or } \CC$ )
with basis $X,Y,Z$ and
 $c_0$ given by $c_0(X,Y) = Z$ and other terms vanishing; so $ {\cal G}_0$ 
 is the 3 dimensional Heisenberg algebra. Let $c_1$ be given by 
$c_1 (X, Z)=X,\ c_1(Y,Z) = - Y,\ c_1 (X,Y) = 0$; so $ {\cal G}_t$ is 
isomorphic to $s \ell (2,k)$. If  $a_0$ is the derivation 
given by $a_0(X) = X,\ a_0 (Y) = - Y,\ a_0 (Z) = 0$, then 
$[[c_1,a_0]] = c_1$ and so $\dot c_t = [[c_t,a_t]]$ which proves 
that the class given by $c_1$ is killed by the deformation.

Let us consider now $ (N,\{,\})$ and  the local
version of the Richardson-Nijenhuis algebra $A^*_{\rm loc}(N)$ (see 
\cite{[L-W]}). 
So let $c_1 \in A^1_{\rm loc} (N)$ be the Vey cocycle, defining 
the infinitesimal part of the Moyal bracket deformation. One should 
check the 1-cohomology of $ {N}$. It is well known since the first 
work of Lichnerowicz about deformations of Lie algebras of vector 
fields that the group  $H^1( {N},  {N})$ is isomorphic to the first 
De Rham cohomology group of the manifold. This isomorphism works as 
follows: let $\a$ be a closed one form on the manifold, it induces a 
derivation $\wtilde \a$ of $ {N}$ through the formula 
$\wtilde \a (f) = \a (H_f)$. An equivalent way of describing this
space is through its isomorphism with the space of symplectic vector fields
modulo the space of hamiltonian vector fields (i.e. all multivalued 
Hamiltonian functions modulo singlevalued Hamiltonian ones). In that case we 
associate to each symplectic vector field the natural derivation 
$f \to L_x f$. But then it is  straightforward to check that symplectic 
vector fields respect the Vey cocycles using the geometric description 
(see \cite{[Ro2],[L-W]}). So following the above notations, the condition 
$[[c_0,a_0]] = 0$ necessarily implies $[[c_1,a_0]]=0$. Thus the above 
conditions can never be satisfied, and the Vey class will  survive under 
the deformation.

It would be interesting  to generalize those arguments  in order to decide 
whether $[ \wtilde V ]$ has a non trivial image in $H^3(\G ,\R)$ or not. 
The situation can again be described through a diagram:
\[
\begin{array}{ccc}
H^2_I(\G; \G) & \hfl{\theta}{} & H^3 (\G;\R) \\
{\scriptstyle \vert} && {\scriptstyle \vert} \\
{\scriptstyle \vert} && {\scriptstyle \vert} \\
{\scriptstyle \downarrow} & &{\scriptstyle \downarrow} \\
H^2_I(N;N) & \hfl{\theta}{} & H^3(N;\R)
\end{array}
\]
where the vertical dotted arrows indicate the contractions. It is easy to
check that $\t([V])$ is non trivial in $H^3(N;\R)$ (see Section 2.1).
So the above diagram gives some confidence in the non vanishing of the 
corresponding class in $H^3(\G;\R)$, but the proof is unknown to the authors.

\section{The embedding of Virasoro algebra into the
completion of $ \G_q$.}
A natural question to ask is whether the standard
 embedding of vector fields ${\got A} (S^1)$ into pseudodifferential symbols
$\G  (S^1)$ can be in some sense quantized, i.e. whether there exists 
a Lie algebra embedding $I_q: {\got A}(S^1) \to \G _q$ which gives 
back the standard one as $q \to 1$. 

If one tries to use the description of  $\psi DO_q$ as a twisted loop 
algebra (\cite{[K-R]}), one identifies the generators 
of ${\got A}(S^1)$ with terms of the form $ {x^n D}$, ${n\in \Z}$ 
belonging to $A_\s \left [ x,x^{-1}\right ]$ (see the above
notations) but then one has: 
$$x^n D x^m D = x^{n+m} \s^m (D) D$$
and so 
$$ [x^n D, x^m D ]= x^{n+m} \bigl ( \s^m (D)D - \s^n (D)D\bigr ) 
= x^{n+m} \left
[ \bigl ( [m]_q - [n]_q \bigr )D + (q^m - q^n) D^2 \right ].$$
This bracket no longer belongs to the image of ${\got A}(S^1)$; the
space spanned by $ {(x^n D)}_{n \in \Z}$ is ``not closed'' in
physicist's language. 

We will look for an embedding of the following kind:
$$I_q : {\got A}(S^1) \to \what{\G }_q \quad\text{ given by }
\quad I_q (x^{n+1} {\part\over \part x}) = x^n f_q (D_q)$$
into some completion of $\G _q$ with analytic $f_q(D)$ approaching $D$
when $q \to 1$.

Replace $A = \C [D,D^{-1}]]$ by $\what A = \C[\log D, D] [ [D^{-1} ] ]$ 
with the automorphism
\begin{align*}
\s(D) &= qD+1 \\
\s(D^{-1}) &= (qD+1)^{-1} = \sum_{k=1}^\infty (-1)^{k-1} q^{-k} D^{-k} \\
\s(\log D) &= \log(qD+1) = \log D + \log q + 
\sum_{k=1}^\infty {(-1)^{k-1}\over k} q^{-k} D^{-k}
\end{align*}
and then let $\what{\G }_q = \{A\log D +B \mid A,B\in A_\s[x,x^{-1}] \}$  
to be the Lie subalgebra of $ \what A_\s  [x,x^{-1}] $. Set
$$f_q(D) = {\log \bigl(1+D(q-1)\bigr ) \over \log q} \overset{\text{def}}=
\frac1{\log q} \Bigl\{ \log D + \log(q-1) - 
\sum_{k=1}^\infty {1\over k(1-q)^k} D^{-k} \Bigr\} $$
and $ I_q : {\got A}(S^1) \to \what{\G }_q$
given by $I_q (x^{n+1} {\part \over \part x}) = x^n f_q (D)$. So 
$$\left [ I_q \left ( x^{n+1} {\part \over \part x} \right ), 
I_q \left( x^{m+1} {\part \over \part x} \right ) \right ] = 
x^{n+m} \left [ \s^m \bigl(f_q(D) \bigr )
 f_q (D) - \s^n \bigl ( f_q (D) \bigr ) f_q (D)\right ].$$
Since $f_q$ is analytic in the neighbourhood of $D=\infty$, one has 
$$\s^m f_q (D) = f_q \bigl (\s^m (D) \bigr ) = 
f_q \bigl ( q^m D + [m]_q \bigr) = 
{1 \over \log q} \log \left (q^m \bigl ( (q-1) D+ 1 \bigr ) \right ).$$
So
$$f_q \bigl (\s^m(D) \bigr ) - f_q \bigl ( \s^n (D) \bigr ) = {1 \over \log
q} (\log q^m - \log q^n) = (m - n).$$
Finally 
$$\left [ I_q \left(x^{n+1} {\part \over \part x}\right ),
I_q \left ( x^{m+1} {\part \over \part x} \right ) \right ] = 
(m-n) I_q \left ( x^{n+m+1} {\part \over \part x} \right ) = 
I_q \left ( \left [ x^{n+1} {\part \over \part x}, 
x^{m+1} {\part \over \part x} \right ] \right ).$$

Now we can check the behaviour of 
Kac-Radul's 
cocycle through this homomorphism. Using any trace on $\what A$, 
for instance $\Tr=\Tr_D$ extended to the whole algebra by 
$\Tr(D^p(\log D)^k)=0$ if $k>0$, one gets:
\begin{align*}
\psi \Bigl(I_q \bigl(x^{n+1} {\part \over \part x}\bigr ), 
& I_q \bigl ( x^{-n+1} {\part \over \part x} \bigr ) \Bigr ) 
= \Tr \Bigl ( \bigl(1 + \s + \ldots +
\s^{n-1}\bigr ) \s^{-n} \bigl ( f_q (D) \bigr ) f_q (D) \Bigr ) \\
&= \Tr \left ( \sum\limits_{i=0}^{n-1} 
\bigl ( {i-n} + f_q (D) \bigr ) \bigl ( i+ f_q (D) \bigr ) \right ) \\
&= \Tr(1) \sum\limits_{i=0}^{n-1} (i-n)i +  
\Tr(f_q(D))\sum\limits_{i=0}^{n-1} (2i - n) + n\Tr(f_q(D)^2) \\
&={n(n+1)(n-1) \over 6}\Tr(1) + n\Tr(f_q(D)^2 - f_q(D)) .
\end{align*}   
So one recovers, up to a multiple, the Virasoro cocycle, the last term
being a coboundary. Let us denote by
$\widehat{\widehat{\G }}_q$ the central extension of 
$\widehat{\G }_q$, defined through the same cocycle.

\begin{prop}
There exists a Lie algebra homomorphism 
$I_q : {\Vir} \to \widehat{\widehat{\G }}_q$ such that 
$I_q$ tends to the canonical embedding as $q \to 1$.
\end{prop}

\begin{rem}
For the above isomorphism between $\G _q$ and the sine-algebra 
$x \mapsto u_1$ and $\part_q + {1 \over q-1} \mapsto u_2$, the map $I_q$ 
gives: 
$I_q \left ( x^n {\part \over \part x} \right ) = 
{\log [(q-1)u_2] \over \log q}$. So in that context, completion means 
adding $\log u_2$ as a formal supplementary variable.
\end{rem}

\section{$q$-KP and $q$-KdV  hierarchies.}
In this section we define $q$-analogs for the KP and $n$-KdV hierarchies 
and show their ``complete integrability''. The $q$-hierarchies in our 
approach are systems describing commutative flows on the space of 
$q$-pseudodifferential operators. In the classical case the extended 
algebra of integral operators (responsible for the hierarchies) and 
the centrally extended Lie algebra of differential operators are dual 
to each other (they are the components of the Manin triple, see 
\cite{[K-Z]}). It would be interesting to investigate such a 
duality for the $q$-analogs.

\subsection{Classical hierarchies.}
Let $Q$ be a pseudodifferential operator of the 
form\break $Q = \part + u_1(x) \part^{-1} + u_2(x) \part^{-2} + \ldots$

\begin{thm}[see, e.g. \cite{[G-D]}]
 For any $m = 1,2,\ldots$ the system
\be 
{\part Q \over \part t_m} = [Q,(Q^m)_+] \label{(7.1)}
\end{equation}
\par\noindent
 a) defines an evolution on the space of $\{Q\}$
\par\noindent
 b) is Hamiltonian (with respect to so called first and second
Gelfand-Dickey brackets)
\par\noindent 
c) defines commuting flows for different $m$.
\par\noindent 
( c$'$) The corresponding Hamiltonians $H_m(Q) = \int \res (Q^m)$ are in
involution and define infinite number of conserved quantities for each
flow).
\end{thm}

\begin{pf} We prove here the statements a) and c), which
later on are generalized to the case of $q$-analogs.

a) We need to show that the vector $[Q,(Q^m)_+]$ belongs to the tangent
space of $\{Q\}$, i.e.~that its differential part vanishes. Indeed,
$[Q,(Q^m)_+] = - [Q,(Q^m)_-]$, and 
$\deg [Q,(Q^m)_-] = \deg Q + \deg (Q^m)_- - 1 = - 1$, 
i.e.~this is an integral operator.

c) Straightforward simple calculation:
\begin{align*}
{\part H_n(Q) \over \part t_k} &= \int \res {\part Q^n \over \part t_k} =
\int \res \sum_{j=0}^{n} Q^j {\part Q \over \part t_k} Q^{n-j-1} \\
 &= \int \res \sum_{j=0}^{n} \left (Q^j [Q,(Q^k)_+]Q^{n-j-1} \right ) =
\int \res [Q^n,(Q^k)_+] = 0 
\end{align*}
The last identity is due to the ad-invariance of trace: the residue of any
commutator is a full derivative. 
\end{pf}

\begin{rem}
The classical KP equation is the
compatibility equation for the flows\break 
$\displaystyle \{ {\part \psi \over \part t_m} = (Q^m)_+ \psi \}$ 
for $m = 2$ and $3$ (see \cite{[G-D]}):
$${\part (Q^3) _+ \over \part t_2} - {\part (Q^2)_+ \over \part t_3} =
[(Q^3)_+,(Q^2)_+]. \eqno (7.2) $$
This is a system of two equations on the coefficients $\{u_1,u_2 \}$, from
which one function can be excluded.
\end{rem}

\begin{rem}
In the same way the phase space of the $n$-KdV
hierarchy is the set of differential operators $\{ L = \part^n + u_{n-2}
\part^{n-2} + \cdots + u_0 \}$. For any operator $L$ there exists the only
pseudodifferential operator $Q$ such that $Q^n = L$, see \cite{[G-D]}
(notation: $Q = L^{1/n}$). Then the $m^{th}$ flow of the $n$-KdV hierarchy
is the system on the coefficients of $L$ :
$${\part L \over \part t_m} = [L,(L^{m/n})_+]. \eqno (7.3)$$
\end{rem}

\begin{thm}[see \cite{[Di],[G-D]}] 
This system
\par\noindent  a) is well-defined
\par\noindent b) is Hamiltonian
\par\noindent
 c) has infinite number of conserved quantities $H_k(L) = \int
\res(L^{k/n})$.
\end{thm}

\begin{pf} To check a) note that by definition
$[L,(L^{m/n})_+] = - [L,(L^{m/n})_-]$, and\break 
$\deg [L,(L^{m/n})_-] = \deg L + \deg (L^{m/n})_- - 1 = n - 1 - 1 = n - 2$, 
i.e.~this is an element of the tangent space of $\{L\}$.

The statement c) follows from theorem above and the fact that $n$-KdV 
flows are the KP flows restricted to the set of those operators $\{Q\}$, 
whose $n^{th}$ power is a purely differential operator:
$${\part L \over \part t_m} = {\part Q^n \over \part t_m} = [Q^n, (Q^m)_+ ]
= [L,(L^{m/n})_+].$$
\end{pf}

\begin{rem}
The classical KdV equation is the first nontrivial
equation in the hierarchy on the space $\{ L = \part^2 + u(x)\}$ (here $n =
2,\ m = 3)$.
\end{rem}

\subsection{$\bold q$-hierarchies.}
The consideration above
almost literally can be applied to the associative algebra of
$q$-pseudodifferential symbols. The phase space for a $q$-KP hierarchy 
is the set $\{Q_q = 
\part_q + u_0(x) + u_1(x) \part_q^{-1} + u_2(x) \part^{-2}_q + \ldots \}$,
and the corresponding system has the same form 
$$ {\part Q_q \over \part t_m} = [Q_q,(Q^m_q)_+] \eqno (7.4)$$
where $+$ means taking purely differential part of $q$-pseudodifferential
operators.  This is a mixed system of
differential equations on the coefficients of $\part^{j}_q$ ($q$-difference
operators).

\begin{thm}
The system $(7.4)$ for any $m$ defines an
evolution on the space $\{Q_q\}$. For any $m$ there is an infinite number
of conservation laws $H^{(q)}_m (Q_q) = \int \res (Q_q^{m}E)$ and the flows
commute for different $m$.
\end{thm}

\begin{pf} Likewise $[Q_q,(Q^m_q)_+] = - [Q_q,(Q^m_q)_-]$, and thus
$\deg {\part Q_q \over \part t_m} = \deg Q_q + \deg (Q^m_q)_- = 1 - 1 =
0$. Now we do not have the cancellation of the leading term in the
commutator of $q$-symbols. This is why we consider here the space of
operators $\{Q_q\}$ containing arbitrary subleading terms $u_0(x)$. The
same calculation as in the classical case verifies the invariance of
$H^{(q)}_m (Q_q)$. We use the {\it ad}-invariance of the trace $\Tr$ for
$q$-operators. 
\end{pf}

\begin{rem}
The analogous $q$-KP equation is the
differential system 
$$ {\part (Q^3_q)_+ \over \part t_2} - {\part(Q_q^2)_+ \over \part t_3} =
[(Q^3_q)_+,(Q_q^2)_+]. \eqno (7.5)$$
It is an overdetermined system, consisting, generally speaking, of 5
equations\break $(\deg \left[(Q_q^3)_+,\ (Q_q^3)_+ \right ] = 5)$ on 3 
unknown coefficients $\{ u_0,u_1,u_2\}$ of $Q_q$. The existence of a 
solution is guaranteed by the Lax formulation above.
\end{rem}

To consider a $q$-analog of the $n$-KdV hierarchy we need to define the
$n^{th}$ root $Q_q$ of a differential operator $L_q = \part^n_q + u_{n-1}
\part_q^{n-1} + \cdots + u_0$. However, instead of uniqueness of the
exponential map of the ``unipotent" group $G$, we have now a sort of
``solvable" group $G_q$ of $q$-operators with the surjective 
but not one-to-one ``exponent". This means that we
have a freedom in the choice of the $n^{th}$-root $Q_q$. 

Then likewise the $m^{th}$ flow of $n^{th}$ KdV hierarchy
$${\part L_q \over \part t_m} = [L_q, (L_q^{m/n})_+ ]$$
for different $m$ the flows commute, and the conserved quantities are
$H_k(L_q) =\break \int \res(L^{k/n} E) = \Tr (L^{k/n})$. Proof repeats the
classical case.

\bibliographystyle{amsplain}

\begin{thebibliography}{99}
\fontsize{10}{12pt}\selectfont

\bibitem{[B-K-K]} {\sc I. Bakas, B. Khesin and E. Kiritsis}, The logarithm 
of the derivative operator and higher spin algebras of $W_{\infty}$ type,
{\em  Commun. Math. Phys.} {\bf151} (1993), 233--243.

\smallskip

\bibitem{[Di]}{\sc L. A. Dickey}, ``Soliton equations and Hamiltonian  
systems,'' Adv. Ser. Math. Phys. {\bf12}, World. Sci., River Edge, NJ, 1991.

\smallskip

\bibitem{Dzha} {\sc A. S. Dhumaduldaev}, Derivations and central
extensions of the Lie algebras of pseudodifferential symbols,
{\em Algebra i Analiz} {\bf 6} (1994) n.~1, 140--158.

\smallskip

\bibitem{[Dr]}{\sc V. G. Drinfeld},  
Quantum groups, {\em in} ``Proceedings of the ICM,'' AMS, 
Providence, R.I. {Vol.~1} (1987),  798--820.

\smallskip

\bibitem{[F-F-Z1]} {\sc D. B. Fairlie, P. Fletcher and C. K. Zachos}, 
Trigonometric structure constants for new infinite-dimensional algebras, 
{\em Phys. Lett.} B {\bf218} (1989), 203--206.

\smallskip

\bibitem{[F-F-Z2]}{\sc  D. B. Fairlie, P. Fletcher and C. K. Zachos}, 
Infinite-dimensional Algebras and a Trigonometric Basis for the Classical
Lie Algebras, {\em J. Math. Phys.} {\bf 31} (1990) n.~5, 1088--1094.

\smallskip

\bibitem{[F-Z]} {\sc D. B. Fairlie and C. K. Zachos}, 
Infinite-dimensional Algebras, Sine Brackets, and $SU(\infty)$,
{\em Phys. Lett.} B {\bf224} (1989) n.~1--2, 101--107.

\smallskip

\bibitem{[Fe]}{\sc B. L. Feigin}, The Lie algebras $gl(\lambda)$ and
cohomologies of Lie algebras of differential operators, 
{\em Russian Math. Surveys} {\bf43} (1988) n.~2, 169--170.

\smallskip

\bibitem{[Fu]}{\sc D. B. Fuchs}, ``Cohomology of infinite-dimensional Lie 
algebras,'' Nauka, 1984.

\smallskip

\bibitem{[G-D]}{\sc I. M. Gelfand and L. A. Dickii}, 
Fractional powers of operators, and Hamiltonian systems, 
{\em Funct. Anal. Appl.} {\bf10} (1976), n.~4, 13--29.



\smallskip

\bibitem{[G-Fu]}{\sc I. M. Gelfand and D. B. Fuks}, 
The cohomologies of the Lie algebra of the vector fields in a circle, 
{\em Funct. Anal. Appl.} {\bf2} (1968) n.~4, 342--343.

\smallskip

\bibitem{[G-M]}{\sc I. M. Gelfand and O. Mathieu}, 
On the cohomology of the Lie algebra of Hamiltonian vector fields, 
{\em J. Funct. Anal.} {\bf108} (1992), 347--360.

\smallskip

\bibitem{[G-S]}{\sc M. Gerstenhaber and S. D. Schack}, Algebraic 
cohomology and deformation theory, {\em in} ``Deformation Theory of Algebras 
and Structures and Applications,'' Nato ASI Series {\bf 247}, \S 7.

\smallskip

\bibitem{[Ge]}{\sc E. Getzler}, 
Cyclic homology and Beilinson--Manin--Schechtman central extension, 
{\em Proc. Amer. Math. Soc.} {\bf104}  (1988) n.~3, 729--734.

\smallskip

\bibitem{[G-L]}{\sc M. I. Golenishcheva-Kutuzova and D. R. Lebedev}, 
$\Z$-graded trigonometric Lie subalgebras in $\hat A_\infty$, 
$\hat B_\infty$, $\hat C_\infty$ and $\hat D_\infty$ and their vertex
operator representations, 
{\em Funct. Anal. Appl.} {\bf27} (1993) n.~1, 10--20.

\smallskip

\bibitem{[Gol]}{\sc M. I. Golenishcheva-Kutuzova}, private communication.

\smallskip

\bibitem{GutdeWil} {\sc S. Gutt, M. Cahen and M. de Wilde},
Local cohomology of the algebra of $C^\infty$ functions on a 
connected manifold, {\em Lett. Math. Phys.} {\bf 4} (1980) n.~3,
157--167.


\smallskip

\bibitem{[H-O-T]}{\sc J. Hoppe, M. Olshanetsky and S. Theisen},
Dynamical systems on quantum tori Lie algebras, 
{\em Commun. Math. Phys.} {\bf155} (1993) n.~3, 429--448.

\smallskip

\bibitem{[Ju]}{\sc D. V. Juriev}, Infinite dimensional geometry
and quantum field theory of strings III, preprint ENS, Paris.

\smallskip

\bibitem{[K-P]}{\sc V. G. Kac and D. M. Peterson}, Spin and wedge 
representations of infinite dimensional Lie algebras and groups,
{\em Proc. Nat. Acad. Sci. USA}, {\bf78} (1981), 3308--3312.

\smallskip

\bibitem{[K-R]}{\sc V. G. Kac and A. O. Radul}, Quasifinite highest weight
modules over the Lie algebra of differential operators on the circle,
{\em Commun. Math. Phys.} {\bf157} (1993) n.~3, 429--457.

\smallskip

\bibitem{[Ka]}{\sc C. Kassel}, Cyclic homology of differential operators, 
the Virasoro algebra and a $q$-analogue, 
{\em Commun. Math. Phys.} {\bf146} (1992), 343--356.

\smallskip

\bibitem{[K-Z]}{\sc B. A. Khesin and I. S. Zakharevich}, Poisson-Lie 
group of pseudodifferential operators and fractional KP-KdV hierarchies,
{\em C. R. Acad. Sci.} {\bf315} S\'er.~I  (1993), 621--626. 
Poisson-Lie group of pseudodifferential operators, preprint IHES/M/93/53, 
arXiv:hep-th/9312088, pp.66.

\smallskip

\bibitem{[Ki]}{\sc A. A. Kirillov}, La g\'eom\'etrie des moments pour les 
groupes de diff\'eomorphismes, {\em in} Progress in Math., 
ed. A.Connes et al {\bf92} (1990), 73--82.

\smallskip

\bibitem{KonVish} {\sc M. Kontsevich and S. Vishik},
Determinants of elliptic pseudo-differential operators. Preprint.

\smallskip

\bibitem{[K-K]}{\sc O. S. Kravchenko and B. A. Khesin}, A nontrivial central
extension of the Lie algebra of pseudodifferential symbols on the circle,
{\em Funct. Anal. Appl.} {\bf25} (1991) n.~2, 83--85.

\smallskip

\bibitem{[Le]}{\sc P. Lecomte}, Applications of the cohomology of graded 
Lie algebras to formal deformations of Lie algebras, 
{\em Lett. Math. Phys.} {\bf13} (1987), 157--166.

\smallskip

\bibitem{[L-R1]}{\sc P. Lecomte and C. Roger}, Central extensions of the Lie 
algebras of Hamiltonian vector fields (in preparation).

\smallskip

\bibitem{[L-R2]}{\sc P. Lecomte and C. Roger}, 
Modules et cohomologie des big\`ebres de Lie, 
{\em C. R. Acad. Sci. Paris} {\bf310} S\'erie I (1990), 405--410.

\smallskip

\bibitem{[L-W]}{\sc P. Lecomte and M. De Wilde}, Formal deformations of the
Poisson-Lie algebra of a symplectic manifold and star products. Existence,
equivalence, derivations, {\em in} ``Deformations theory of algebras and
structures,'' NATO-ASI Series C 297, Kluwer Academic Publishers.

\smallskip

\bibitem{[Li]}{\sc W.-L. Li}, 2-cocycles on the Lie algebra of 
differential operators, {\em J. Algebra}, {\bf122} (1989), 64--80.

\smallskip

\bibitem{[M-M-R]}{\sc F. Martinez Moras, J. Mas and E. Ramos}, 
Diffeomorphisms from higher dimensional $W$-algebras, 
{\em Modern Phys. Lett.} A {\bf 8} (1993) n.~23, 2189--2197.

\smallskip

\bibitem{[P-R-S]}{\sc C. N. Pope, L. J. Romans and X. Shen}, $W_\infty$ 
and the Racah-Wigner algebra, {\em Nuclear Phys.} B{\bf339} (1990), 191--221.

\smallskip

\bibitem{[P-S]}{\sc A. Pressley and G. Segal}, ``Loop groups,'' 
Cambridge University Press, 1984.

\smallskip

\bibitem{[Ro1]}{\sc C. Roger}, Alg\`ebres de Lie gradu\'ees et 
quantification, {\em in} ``Symplectic geometry and Mathematical Physics,"
Progress in Math. {\bf99}, Birkh\"auser-Verlag, 1991.

\smallskip

\bibitem{[Ro2]}{\sc C. Roger}, Extensions centrales d'alg\`ebres et de 
groupes de Lie de dimension infinie, alg\`ebre de Virasoro et 
generalisations, Preprint of Inst. de Math., Univ. de Liege, 1993.

\smallskip

\bibitem{[S_V]}{\sc M. V. Saveliev and A. M. Vershik}, 
New examples of continuum graded Lie algebras, 
{\em Phys. Lett.} {\bf A143} (1990) n.~3, 121--128.

\smallskip

\bibitem{[Ve]} {\sc J. Vey}, 
Deformation du crochet de Poisson sur une variet\'e symplectique, 
{\em Comment. Math. Helv.} {\bf 50} (1975) n.~4, 421--454.

\smallskip

\bibitem{Wam:pDO} {\sc M. Wambst},
Homologie de l'alg\`ebre quantique des symboles pseudo-diff\'erentiels
sur le cercle. Preprint.

\smallskip

\bibitem{[Wo]} {\sc M. Wodzicki}, Cyclic homology of differential 
operators, {\em Duke Math. J.} {\bf54} (1987), 641--647.

\end{thebibliography}


\end{document}